\documentclass[apj,iop]{emulateapj}
\usepackage{graphicx,natbib}
\newcommand{\lsun}{L_{\odot}}
\newcommand{\lbol}{L_{\rm bol}}
\newcommand{\lir}{L_{\rm IR}}

\newcommand{\msun}{M_{\odot}}
\newcommand{\msunperyr}{M_{\odot} {\rm ~yr}^{-1}}

\newcommand{\mdust}{M_{d}}
\newcommand{\tdust}{T_{d}}
\newcommand{\ldust}{L_{d}}

\newcommand{\sunrise}{\textsc{Sunrise}~}

\newcommand{\gadgettwo}{\textsc{Gadget-2}~}

\begin{document}

\title{What does a submillimeter galaxy selection actually select? \\ The dependence of submillimeter flux density on \\ star formation rate and dust mass}
\author{Christopher C. Hayward\altaffilmark{1,2}, Du\v{s}an Kere\v{s}\altaffilmark{1,3,4}, Patrik Jonsson\altaffilmark{1},\\
Desika Narayanan\altaffilmark{1,5,6}, T. J. Cox\altaffilmark{7}, and Lars Hernquist\altaffilmark{1}}
\altaffiltext{1}{Harvard-Smithsonian Center for Astrophysics, 60 Garden Street, Cambridge, MA 02138, USA}
\altaffiltext{2}{chayward@cfa.harvard.edu}
\altaffiltext{3}{Department of Astronomy and Theoretical Astrophysics Center, University of California Berkeley, Berkeley, CA 94720, USA}
\altaffiltext{4}{Hubble Fellow}
\altaffiltext{5}{Steward Observatory, Department of Astronomy, University of Arizona, 933 North Cherry Avenue, Tucson, AZ 85721, USA}
\altaffiltext{6}{Bart J. Bok Fellow}
\altaffiltext{7}{Carnegie Observatories, 813 Santa Barbara Street, Pasadena, CA 91101, USA}
\shorttitle{Dependence of Submm Flux on SFR \& Dust Mass}
\shortauthors{C.~C. Hayward et al.}

%\date{Accepted ???. Received ???; in original form ???}

%\pagerange{\pageref{firstpage}--\pageref{lastpage}} \pubyear{2010}

\begin{abstract}

We perform 3-D dust radiative transfer (RT) calculations on hydrodynamic simulations of isolated and merging disk galaxies
in order to quantitatively study the dependence of observed-frame submillimeter (submm) flux density on galaxy properties.
We find that submm flux density and star formation rate (SFR) are related in dramatically different ways for quiescently star-forming
galaxies and starbursts. Because the stars formed in the merger-induced starburst do not dominate the bolometric luminosity and
the rapid drop in dust mass and more compact geometry cause a sharp increase in dust temperature during the burst,
starbursts are very inefficient at boosting submm flux density (e.g., a $\ga16$x boost in SFR yields a $\la 2$x boost in submm flux density).
Moreover, the ratio of submm flux density to SFR differs significantly between the two modes; 
thus one cannot assume that the galaxies with highest submm flux density are necessarily those with the highest bolometric luminosity or SFR.
These results have important consequences for the bright submillimeter-selected galaxy (SMG) population.
Among them are: 1. The SMG population is heterogeneous. In addition to merger-driven starbursts,
there is a subpopulation of galaxy pairs, where two disks undergoing a major merger but not yet strongly interacting are blended into one submm
source because of the large ($\ga 15$'', or $\sim 130$ kpc at $z = 2$) beam of single-dish submm telescopes. 2. SMGs must be very massive
($M_{\star} \ga 6 \times 10^{10} \msun$). 3. The infall phase makes the SMG duty cycle a factor of a few greater than what is expected
for a merger-driven starburst. Finally, we provide fitting functions for SCUBA and AzTEC submm flux densities as a function
of SFR and dust mass and bolometric luminosity and dust mass; these should be useful for calculating submm flux density in semi-analytic models
and cosmological simulations when performing full RT is computationally not feasible.

\end{abstract}

% XXX add keyword(s) for dust, IR, SED fitting... (if I can have more)

\section{Introduction}

Submillimeter-selected galaxies (SMGs; \citealt{Smail:1997,Barger:1998,Hughes:1998,Eales:1999}; see \citealt{Blain:2002} for a review)
are extremely luminous \citep[bolometric luminosity $\lbol \sim 10^{12} - 10^{13} L_{\odot}$; e.g.,][]{Kovacs:2006},
high-redshift \citep{Chapman:2005} galaxies powered primarily by star formation rather than AGN
\citep{Alexander:2005,Alexander:2005b,Alexander:2008,Valiante:2007,Menendez:2007,Menendez:2009,Pope:2008MIR,
Younger:2008phys_scale,Younger:2009EGS}.
Because of their high dust content, SMGs emit almost all of their luminosity in the IR.
As the name suggests, a galaxy is defined as an SMG if it is detected in the submm (historically, 850 \micron ~flux density $S_{850}
\ga 3-5$ mJy; the nature of the population is sensitive to the adopted flux density cut, so we define an SMG as a source with $S_{850} > 3$ mJy),
which requires $\lir \ga 10^{12} \lsun$ \citep{Kovacs:2006,Coppin:2008}, so SMGs are typically ultra-luminous infrared galaxies (ULIRGs).
Locally, ULIRGs are almost exclusively merging galaxies \citep{Sanders:1996,Lonsdale:2006}, so one might expect that 
at least some SMGs are also merging galaxies. Indeed, many observations support a merger origin for SMGs
\citep[e.g.,][]{Ivison:2002,Ivison:2007,Ivison:2010,Chapman:2003, Neri:2003,Smail:2004,Swinbank:2004,Greve:2005,Tacconi:2006,Tacconi:2008,
Bouche:2007,Biggs:2008,Capak:2008,Younger:2008phys_scale,Younger:2010,Iono:2009,Engel:2010}.
Furthermore, in \citet[][hereafter N10]{Narayanan:2010smg} we combined hydrodynamic simulations
and radiative transfer (RT) calculations to show that major mergers
can reproduce the full range of submm flux densities and typical UV-mm spectral energy distribution (SED) of SMGs (cf. \citealt{Chakrabarti:2008SMG};
\citealt{Chakrabarti:2009}). Semi-analytic models also predict that the SMG population is dominated by merger-induced starbursts rather than quiescent star formation
(\citealt{Baugh:2005,Fontanot:2007,Swinbank:2008,LoFaro:2009,Fontanot:2010,Gonzalez:2011}; but cf. \citealt{Granato:2004}).

However, because of the much greater rate of gas supply onto galaxies at high redshift
\citep[e.g.,][]{Keres:2005,Dekel:2009nature}, gas fractions \citep{Erb:2006,Tacconi:2006,Tacconi:2010,Daddi:2010} and star formation rates
\citep{Daddi:2007,Noeske:2007b,Noeske:2007a} of galaxies at fixed galaxy mass increase rapidly with redshift.
Thus, at $z \sim 2-3$ even a ``normal'' star-forming galaxy can reach ULIRG luminosities
\citep[e.g.,][]{Hopkins:2008cosm_frame1,Hopkins:2010IR_LF,Daddi:2005,Daddi:2007,Dannerbauer:2009}.
Furthermore, roughly estimating submm counts using estimates of high-redshift major merger rates and the short duty cycle of merger-induced
starbursts suggests that there may not be enough major mergers to account for the SMG population \citep{Dave:2010}. This motivates the view
that, instead, typical SMGs may be massive, gas-rich disks quiescently forming stars and fueled by continuous gas supply from mergers and
smooth accretion (\citealt{Carilli:2010}, but cf. \citealt{Daddi:2009smgb}).

The mode of star formation responsible for the majority of the SMG population is still a matter of debate, as it is difficult
to discriminate between the two scenarios given the currently available data.
A better understanding of the submm galaxy selection can clarify the nature of the SMG population.

Since SMGs have redshifts
$z \sim 1-4$ \citep{Dannerbauer:2002,Chapman:2005,Younger:2007high-z_SMGs,Younger:2008phys_scale,Capak:2008,Greve:2008,Schinnerer:2008,
Daddi:2009smga,Daddi:2009smgb,Knudsen:2010},
the observed submm flux density traces rest-frame $\sim 150 - 400$ \micron, longward of the peak of the IR SED.
Thus the observed submm flux density is sensitive to both the total IR luminosity and the ``dust temperature''\footnote{As
is convention, we will use the term ``dust temperature'' to denote the temperature derived from a single-temperature modified blackbody fit
to the SED. This is simply a parameterization of the SED shape rather than a physical temperature. In our simulations dust grains
have a continuum of temperatures, depending on both grain size and the local radiation field heating the dust.}
of the SED, which depend on the luminosity from stars and AGN absorbed by the dust, the mass and composition of the dust,
and the spatial distribution of stars, AGN, and dust. Galaxies do not have identical SED shapes, so the dependence on dust temperature implies
that galaxies with the highest submm flux density are not necessarily those with the highest bolometric luminosity. Furthermore, because star formation
histories are more complicated than an instantaneous burst, the luminosity and instantaneous SFR are not necessarily linearly proportional.
Thus the relationship between submm flux density and SFR is potentially more complicated than the relationship between submm flux density and bolometric
luminosity. We therefore cannot say \emph{a priori} that the
galaxies with the highest submm flux densities are the most rapidly star-forming or most luminous bolometrically. Indeed, it has already
been observationally demonstrated that submm selection does not select all the brightest galaxies in a
given volume, as there are galaxies with luminosities and redshifts comparable to those of SMGs that are undetected in the
submm because of their relatively hot SEDs \citep{Chapman:2004,Chapman:2010,Casey:2009,Casey:2010,Hwang:2010dust_T_evolution,
Magdis:2010dust_T,Magnelli:2010}.
A submm galaxy selection is clearly biased toward cold galaxies; however, the details of the selection bias are yet to be understood.

Despite the basic physical reasons that one does not expect a simple relation between submm flux density and SFR, a linear relation
between submm flux and SFR has been used explicitly (and, even more frequently, implicitly) to infer SFR from observed submm flux densities
\citep[e.g.,][]{Chapman:2000,Peacock:2000,Blain:2002,Scott:2002,Webb:2003,vanKampen:2005,Tacconi:2008,Wang:2011}, typically because
the data
sets do not have enough photometric data points to precisely constrain the IR SED shape (\emph{Herschel} data are already helping greatly in this regard;
e.g., \citealt{Chapman:2010,Dannerbauer:2010,Magnelli:2010}).
Furthermore, some theoretical studies \citep{Dave:2010} have assumed that SMGs are the most rapidly star-forming galaxies
in order to identify SMGs in cosmological simulations without performing RT. If SFR and submm flux density are not simply related this approach is problematic.

It is clear that a better understanding of the relationship between submm flux density and SFR and, more generally, what galaxy properties a
submm galaxy selection selects for, is needed.
In other work we have combined hydrodynamic simulations and dust RT to show that major mergers
of massive, gas-rich disk galaxies can reproduce the 850 \micron ~flux densities (N10),
CO properties \citep{Narayanan:2009}, number densities \citep[][C. Hayward et al.~2011, in preparation]{Hayward:2011num_cts_proc},
and intersection with the dust-obscured galaxy (DOG) population
\citep{Narayanan:2010dog} of SMGs. Motivated by the success of our simulations in reproducing a variety of SMG properties,
we use them here to quantify how submm flux density depends on SFR, $\lbol$, dust content, and geometry.
The aim of this study is to clarify for what galaxy properties a submm selection criterion selects and to provide a discriminant among the different
modes of star formation that could power SMGs.

\section{Methods}

We combine high-resolution \gadgettwo \citep{Springel:2001gadget,Springel:2005gadget}
3-D N-body/smoothed-particle hydrodynamics (SPH) simulations with the \sunrise \citep{Jonsson:2006sunrise,Jonsson:2010sunrise} polychromatic Monte Carlo
dust RT code in order to predict the submillimeter flux densities of high-redshift isolated and merging disk galaxies.
The simulations presented here, part of a larger suite to be presented in C. Hayward et al. (2011, in preparation),
are described in N10, so here we will only summarize and describe differences from that work.
This combination of \gadgettwo and \sunrise has been successfully shown to reproduce the SEDs/colors of local SINGS \citep{Kennicutt:2003,Dale:2007}
galaxies \citep{Jonsson:2010sunrise}; local ULIRGs \citep{Younger:2009}; massive, quiescent, compact $z \sim 2$ galaxies \citep{Wuyts:2009b,Wuyts:2010}; 
24 \micron-selected galaxies \citep{Narayanan:2010dog}; K+A/post-starburst galaxies \citep{Snyder:2011}; and
XUV disks \citep{Bush:2010}, among other populations.
The success of our approach at modeling diverse galaxy populations---both local and high-redshift---lends credibility to its application to modeling SMGs.

\subsection{Hydrodynamic simulations}

{\sc Gadget-2}\footnote{A public version of \gadgettwo is available at \url{http://www.mpa-garching.mpg.de/$\sim$volker/gadget/index.html}.} % XXX URL
\citep{Springel:2001gadget,Springel:2005gadget} is a TreeSPH \citep{Hernquist:1989treesph} code that computes gravitational interactions via a
hierarchical tree method \citep{Barnes:1986} and gas dynamics via SPH \citep{Lucy:1977,Gingold:1977}. It conserves both energy and entropy \citep{Springel:2002}.
The simulations include radiative heating and cooling as in \citet{Katz:1996}. Star formation is modeled via the volume-density
dependent Kennicutt-Schmidt
law \citep{Kennicutt:1998}, $\rho_{\rm SFR} \propto \rho_{\rm gas}^{1.5}$, with a minimum density threshold; this index is consistent
with observations of $z \sim 2$ disks \citep{Krumholz:2007KS,Narayanan:2008CO_SFR,Narayanan:2011ks}.
The density threshold used is $n \sim 0.1$ cm$^{-3}$, much less than that of the dense molecular gas from which stars form ($n \sim 10^2 - 10^3$ cm$^{-3}$).
For this reason, and because we do not track the formation of molecular gas, the KS law employed should be considered simply an empirically- and
physically-motivated prescription to encapsulate physics we do not resolve. The SF prescription has been calibrated to reproduce the global K-S law
(see \S3 of \citealt{Springel:2003}).
Recently, some authors \citep[e.g.,][]{Hopkins:2011self-regulated_SF} have presented simulations that resolve the density threshold for molecular gas formation;
we plan to compare such simulations to our current simulations in future work.

The structure of the ISM is modeled via a two-phase sub-resolution model in which cold, dense clouds are embedded
in a diffuse, hot medium \citep{Springel:2003}. This medium is pressurized by
supernova feedback, which heats the diffuse ISM and evaporates the cold clouds \citep{Cox:2006feedback}. Metal enrichment is calculated by assuming
each gas particle behaves as a closed box.
Black hole particles accrete via Eddington-limited Bondi-Hoyle accretion and deposit 5\% of their emitted luminosity---calculated
from the accretion rate assuming 10\% radiative efficiency, $\lbol = 0.1 \dot{M}c^2$---to the surrounding
ISM as thermal energy \citep{Springel:2005feedback,DiMatteo:2005}. We refer the reader to the references given above for the full
details of the \gadgettwo code and the sub-resolution models employed.

We focus on two simulations,
one isolated disk and one major merger. We embed exponential disks with baryonic mass
$4 \times 10^{11} M_{\odot}$ in $9 \times 10^{12} M_{\odot}$ dark matter halos described by a
\citet{Hernquist:1990} density profile. The disks are initially 60\% gas and
are scaled to $ z \sim 3$ as described in \citet{Robertson:2006disk_formation,Robertson:2006}.
The gravitational softening lengths are 200$h^{-1}$ pc for the dark matter particles and 100$h^{-1}$ pc for the star, gas,
and black hole particles. We use $6 \times 10^4$ dark matter, $4 \times 10^4$ stellar, $4 \times 10^4$
gas, and 1 black hole particle per disk galaxy. For the major merger, we initialize two such disks on parabolic orbits
with initial separation $R_{\rm init} = 5R_{\rm vir}/8$ and pericentric distance twice the disk scale length \citep{Robertson:2006}.
The orbit we focus on is the `e' orbit of \citet{Cox:2006}. We have checked that the differences between quiescent star formation and
starbursts are insensitive to orbit as long as a strong starburst is induced (some orbits do not induce strong starbursts, but those are irrelevant
for the purpose of studying the differences between starbursts and quiescent star formation), and the larger suite of simulations
used to derive the fitting functions includes a variety of orbits.

\subsection{Radiative transfer}

Every 10 Myr we save snapshots of the \gadgettwo simulations and use the 3-D Monte Carlo
dust RT code {\sc Sunrise}\footnote{{\sc Sunrise} is publicly available at \url{http://code.google.com/p/sunrise/}.}
\citep{Jonsson:2006sunrise,Jonsson:2010sunrise} in post-processing to
calculate the SEDs of the simulated galaxies. While we will summarize the key features of
\sunrise here, we encourage the reader to see \citet{Jonsson:2006sunrise} and \citet{Jonsson:2010sunrise}
for full details. Except where noted, we use the fiducial parameters given in \citet{Jonsson:2010sunrise}.
{\sc Sunrise} calculates the emission from the stars
and AGN in the \gadgettwo simulations and the attenuation and re-emission from dust. 
{\sc Starburst99} \citep{Leitherer:1999} SEDs are assigned to all star particles according to their
ages and metallicities. Star particles present at the start of the \gadgettwo simulation are assigned ages
assuming that their stellar mass was formed at a constant rate equal to the star formation rate of the
initial snapshot and gas and stellar metallicities via a closed-box model, $Z = -y \ln f_g$, where $f_g$ is the initial
gas fraction and $y = 0.02$. Black hole particles are assigned the luminosity-dependent templates of
\citet{Hopkins:2007} by assuming that the bolometric luminosity of a black hole particle is $\lbol = 0.1 \dot{M} c^2,$
where $\dot{M}$ is the black hole accretion rate from the \gadgettwo simulations.

To calculate the dust density, and thus optical depth along a given line-of-sight, \sunrise projects the \gadgettwo
gas-phase metal density
onto a 3-D adaptive mesh refinement grid using the SPH smoothing kernel. We have assumed 40\% of the metals
are in dust \citep{Dwek:1998,James:2002}. We use a maximum refinement level of 10, resulting in a minimum
cell size of 55$h^{-1}$ pc.
By performing runs with higher levels of refinement we have checked that the observed-frame
submm flux density is converged to within 10\%.
We use the Milky Way R=3.1 dust model of \citet{Weingartner:2001} as updated by \citet{Draine:2007}.
Dust models with different FIR opacity will lead to different relationships between submm flux density
and dust mass, but we show how to rescale for different values of the opacity in Equation \ref{eq:S_nu_general}.

Once the dust density grid is constructed and the input sources are assigned SEDs, \sunrise performs
Monte Carlo RT by randomly emitting photon packets from the source particles and randomly drawing
interaction optical depths from the appropriate probability distribution. We use $10^7$
photon packets total for each stage of the RT, having confirmed that this results in Monte Carlo
noise of less than a few percent. The photon packets are scattered and absorbed by dust as they
traverse the ISM. For each grid cell, the temperature of each dust species is calculated assuming
the dust is in thermal equilibrium, and the dust re-emits the absorbed energy as a modified
blackbody.

A \sunrise feature key to this work is its treatment of dust self-absorption. In high-density regions, the
dust can be opaque to its own emission, so the contribution of the dust emission to dust heating
must be computed in addition to the contribution from stars and AGN. {\sc Sunrise} computes the
equilibrium dust temperatures self-consistently by iteratively performing the transfer of the dust emission
and the temperature calculation using a reference field technique similar to that of \citet{Juvela:2005}.
(The details of the \sunrise
implementation are in
\citealt{Jonsson:2010sunrise} and \citealt{Jonsson:2010gpu}.)
This algorithm ensures accurate
dust temperatures, and thus submm flux densities, even for the highly optically thick central starbursts.

The results of the \sunrise calculation are
spatially resolved, multi-wavelength (we use only 120 wavelengths here because of memory constraints)
SEDs observed from 7 cameras distributed isotropically
in solid angle, though in this paper we only utilize the integrated flux densities in the SCUBA \citep{Holland:1999}
and AzTEC \citep{Wilson:2008} bands. For the purpose
of calculating observed flux densities we assume the simulated galaxies are at redshift $z = 2$.

\subsubsection{Differences from Narayanan et al.} \label{S:diff_from_nar}

The primary difference between our simulations and those of N10 and \citet{Narayanan:2010dog}
is the treatment of the ISM on sub-resolution scales.
In order to model the effects of HII and photodissociation regions (PDRs),
\sunrise assigns star particles with ages less than 10 Myr SEDs from the HII region template library of \citet{Groves:2008}.
The time-averaged fraction of solid angle obscured by the PDR, $f_{\rm PDR}$, strongly affects the
resulting attenuation and IR emission \citep[for a detailed discussion see][]{Groves:2008}.
Narayanan et al. assumed $f_{\rm PDR} = 1$ (so that the young stars are
completely obscured by PDRs for 10 Myr) in order to match the observed range
of 850 \micron ~flux densities. Furthermore, they neglected the dust associated with the cold phase of
the \citet{Springel:2003} ISM model, typically $\ga30$\% of the total gas mass
and $\ga 90$\% of the gas mass in the central regions for snapshots classified as SMGs.

Motivated by concerns over applicability of the \citet{Groves:2008} models to the extreme ISM densities and
pressures encountered in our simulations, we set $f_{\rm PDR} = 0$, eliminating all significant
dust obscuration from the sub-resolution PDR model. Instead, we use the total gas density in the SPH simulations
(i.e., both the diffuse and cold phases) to calculate the dust density. Since the dust mass implicit in the \citet{Groves:2008}
PDRs is not tied to the total dust mass of the simulated galaxy, it is possible that one can have more dust mass in the sub-resolution
PDRs than is available in the galaxy. It is also possible that the sum of the dust mass in the PDRs is less than the total available
in the cold phase of the sub-resolution ISM. Our treatment ensures that neither scenario can occur.

Our assumed ISM structure (cold phase volume filling factor of unity) is similar to what is observed for the dense cores of local ULIRGs
\citep{Scoville:1991,Downes:1998,Sakamoto:1999,Sakamoto:2008,Papadopoulos:2010}. Furthermore, it leads to effective far-IR optical depths (inferred from modified
blackbody fitting using $L_{\nu} \propto (1-e^{-\tau_{\nu}}) B_{\nu}(T_d)$; see below for details)
consistent with what is observed for local ULIRGs and SMGs, $\tau = 1$ at rest-frame $\lambda \sim 200 \micron$ for the simulations versus
at rest-frame $\lambda \sim 200-270 ~\micron$ for local ULIRGs \citep{Lisenfeld:2000,Papadopoulos:2010,Rangwala:2011} and SMGs \citep{Lupu:2010,Conley:2011}
(but cf. \citealt{Kovacs:2010}). However, it is still important to note that the sub-resolution ISM structure
is the key uncertainty in these calculations \citep{Younger:2009}.
While unresolved clumpy dust can significantly affect the resulting SED \citep{Witt:1996,Varosi:1999a},
a more detailed treatment is beyond the scope of this work.
The trends presented in this work should be qualitatively insensitive to the sub-resolution ISM assumption
because, as explained  below, the dominant drivers of the differences between the quiescent star
formation and starburst cases are the contribution from stars formed pre-burst to the luminosity at the time of the burst and
the rapid gas consumption during the burst, both of which do not depend on the treatment of sub-resolution
clumpy dust.

\section{Results}

\begin{figure*}[htb]
\epsscale{1.15}
\plottwo{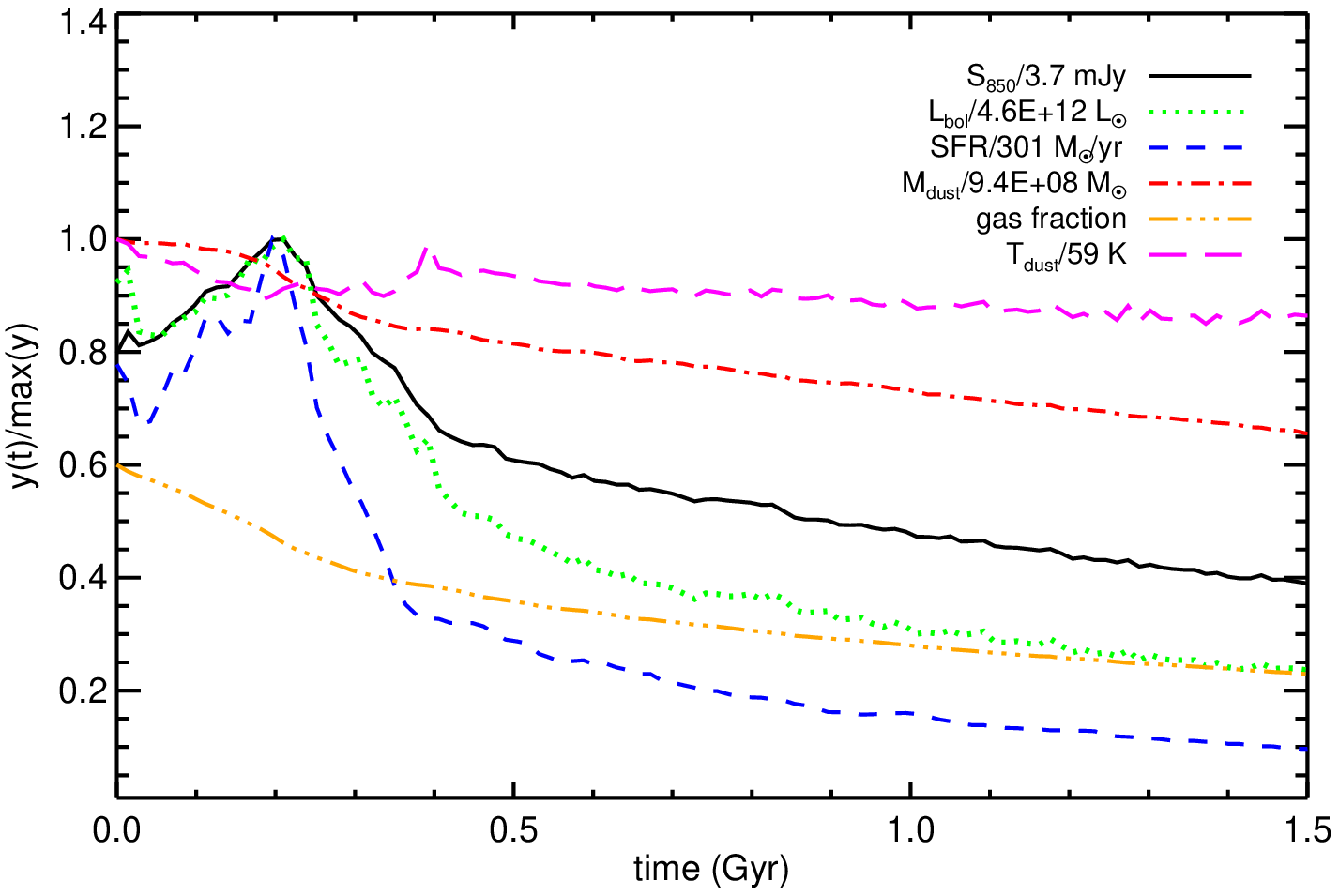}{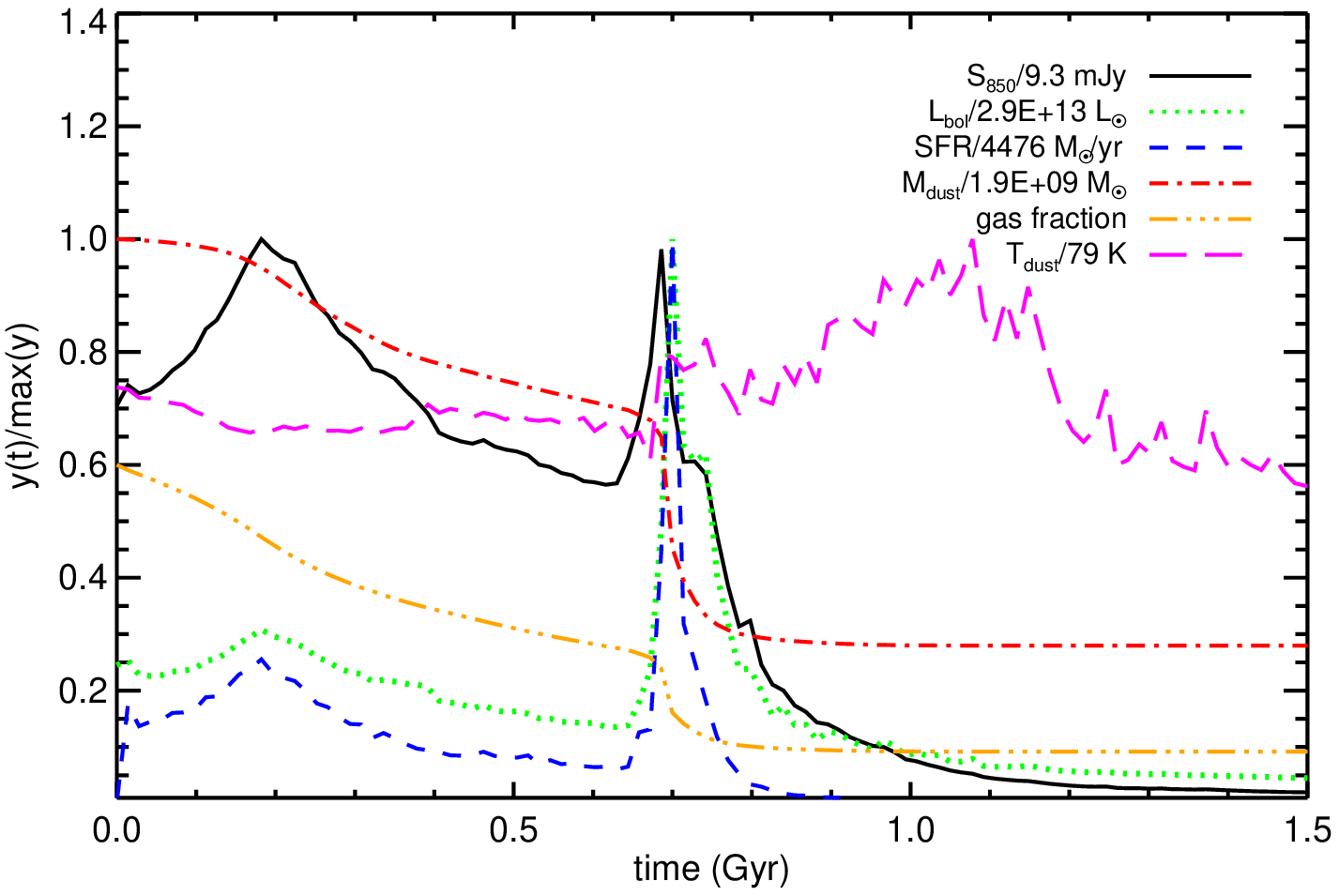}
\caption{\emph{Left:} The isolated disk simulation's observed-frame integrated
SCUBA 850 \micron ~flux density (mJy, assuming $z = 2$; solid black) measured from one of the seven viewing angles, bolometric
luminosity ($\lsun$; dotted green), SFR ($\msunperyr$; dashed blue), dust mass $\mdust$ ($\msun$; dash dot red),
gas fraction $f_g$ (dash dot dot dot orange), and dust temperature (derived from SED fitting) $\tdust$ (K; long dashed magenta)
versus time (Gyr). Except for $f_g$, the quantities have been normalized by dividing by their maximum values,
given in the legend. Once the disk reaches equilibrium, $S_{850}$, $\lbol$, SFR, and $f_g$ concomitantly decrease exponentially with time as the gas
is converted into stars. $\mdust$ also decreases, but by less than a factor of 2, because the decreasing $f_g$ is offset by the increasing
metallicity of the gas. $\tdust$ decreases from $\sim 60$ K to $\sim 50$ K.
\emph{Right:} Same, but for the major merger simulation. All quantities are totals for the two-disk system. Compared to the isolated disk the time evolution
is more complex: Initially $S_{850}$, SFR, $\lbol$, and $\mdust$ are roughly just the sums of the values for the isolated progenitor disks.
At first passage ($t \sim 0.1$ Gyr), the SFR is not elevated much beyond the baseline rate because the disks are very gas-rich and thus lack the massive
stellar bar needed to efficiently remove angular momentum from the gas \citep{Hopkins:2009disk_survival}.
As the two disks coalesce ($t \sim 0.7$ Gyr), tidal torques cause a burst of star formation, resulting in the sharp increase in the SFR ($\ga 16$x), $\lbol$ ($\sim 7$x),
and $S_{850}$ ($\la 2$x)
at $\sim0.7$ Gyr. The increase in $\lbol$ is much less than that boost in SFR because the luminosity of the stars formed during the burst is only $\sim 6$x
the luminosity from stars already formed pre-burst.
$\tdust$ increases sharply from $\sim 50$ K to $\sim 65$ K because of the strong increase in $\lbol$, concurrent decrease in $\mdust$, and more
compact geometry. This mitigates the increase in $S_{850}$ that occurs from increased $\lbol$.
The second, minor peak in $\lbol$, which occurs $\sim 40$ Myr after the peak SFR, corresponds to the peak AGN luminosity. Pre-coalescence,
$f_g$ decreases at a rate similar to the isolated disk case. At coalescence the gas is rapidly consumed in the central starburst.
$\mdust$ decreases by a factor of 4.5, with the bulk of the decrease occurring at coalescence.}
\label{fig:iso_evolution}
\end{figure*}

Figure \ref{fig:iso_evolution} shows the time evolution of the observed SCUBA 850 \micron ~flux density (mJy), SFR ($\msunperyr$;
calculated by dividing the mass of stars formed in the last 10 Myr of the simulation by 10 Myr),
bolometric luminosity $\lbol$ ($\lsun$), initial fraction of baryonic mass that is gas $f_g$, dust mass $\mdust$ ($\msun$), and dust
temperature\footnote{We have calculated $\tdust$ by fitting the modified blackbody $L_{\nu} \propto (1-\exp[-(\nu/\nu_0)^{\beta}])B_{\nu}(\tdust)$ to the rest-frame 20 - 1000
\micron ~SED, allowing all parameters to vary. Here $\nu_0$ is the frequency at which the effective optical depth $\tau_{\nu} = 1$.
We have assumed the opacity has a power-law dependence on $\nu$ in the IR, $\kappa_{\nu} \propto \nu^{\beta}$.
Note we have not used the optically thin form $L_{\nu} \propto \nu^{\beta} B_{\nu}(\tdust)$,
as is almost always done, because we find that the first form, which does not assume optical thinness,
provides a significantly better fit to our SEDs. This is because our simulated SMGs can be optically thick out to
rest-frame $\ga 100 ~\micron$, which is supported by recent \textit{Herschel} observations from
\citet{Lupu:2010} and \citet{Conley:2011}, who found $\tau \sim 1$ at rest-frame $\lambda \sim 200 ~\micron$. Our fitting procedure gives systematically
higher $\tdust$ (by as much as 20 K) than when the optically thin form is used, so comparisons of
our dust temperatures to other results should take this into account.} $\tdust$ (K)
for the isolated disk galaxy (\emph{left}) and merger (\emph{right}), where all quantities except $f_g$ have been normalized by dividing by their
maximum values, given in the legend. When calculating the observed flux density we have assumed the simulated galaxy is at $z = 2$.
The disk is somewhat unstable initially; after the disk settles, $\lbol$, SFR, and $S_{850}$
decrease steadily with time. Over the 2 Gyr of the simulation the gas
fraction decreases from 60\% to 20\%. As the gas is consumed, the SFR, and thus $\lbol$, both decrease,
by factors of $\sim 10$ and $\sim 5$, respectively. $S_{850}$ decreases by $\sim 2.5$x.
$\mdust$ also decreases as a result of the decrease in gas mass,
but only by $\sim 40\%$ because the decrease in gas mass is partially mitigated by metal enrichment of the gas from
star formation, as the metallicity doubles over the course of the simulation.
While it may seem counter-intuitive that $\mdust$ decreases with time, for
a simple closed-box model assuming dust traces metals and constant yield it can be shown \citep{Edmunds:1998}
that for $f_g \la 0.6$ the dust mass increases at most by $\sim 0.1$ dex, and it decreases monotonically with $f_g$
for $f_g \la 0.4$. Furthermore, the preferential consumption of metal-enriched gas that occurs in our models should result
in a lower dust mass than the simple closed-box case of \citet{Edmunds:1998}, which assumes perfect mixing.

The behavior of the merger (Figure \ref{fig:iso_evolution}, right) is qualitatively different from that of the isolated disk. Initially,
SFR, $\lbol$, and $S_{850}$ are roughly equal to the sum of the isolated values for the two progenitor disks,
because the disks are too gas-rich at first passage ($t \sim 0.1$ Gyr) for tidal torques to cause a strong starburst, as a significant stellar bar is
required for the gas to efficiently loose angular momentum \citep{Hopkins:2009disk_survival}.
However, at final coalescence of the two disks ($\sim 0.7$ Gyr) tidal torques induce a strong starburst, causing the SFR
to increase by a factor of $\ga16$.
The peak of the burst is very narrow and significant luminosity from previously formed
stars remains, so the bolometric luminosity increases by a much smaller amount ($\sim 7$x) than the SFR.
Moreover, as the gas is rapidly consumed in the starburst, $\mdust$ plummets by a factor of 3.
Along with the more compact geometry, the decreased dust mass causes the SED to become hotter, with $\tdust$ increasing from $\sim 50$ K to $\sim 65$
K.
The increase in dust temperature during the starburst is qualitatively consistent with observations, as local ULIRGs (i.e., merger-induced starbursts)
tend to have hotter dust temperatures ($\sim 42$ K) than less luminous (quiescent) galaxies ($\sim 35$ K) \citep{Clements:2010}.
The increased $\tdust$ partially offsets the increase in $S_{850}$ caused by the increased luminosity. The combination
of the significant pre-burst contribution to $\lbol$, the small mass of stars formed in the burst,
and the increased $\tdust$ cause $S_{850}$ to increase by $\la 2$x even though the SFR increases
by $\ga 16$x in the burst.

\subsection{The relationship between submm flux density and SFR} \label{S:S850-SFR}

\begin{figure*}
\epsscale{1.15}
\plottwo{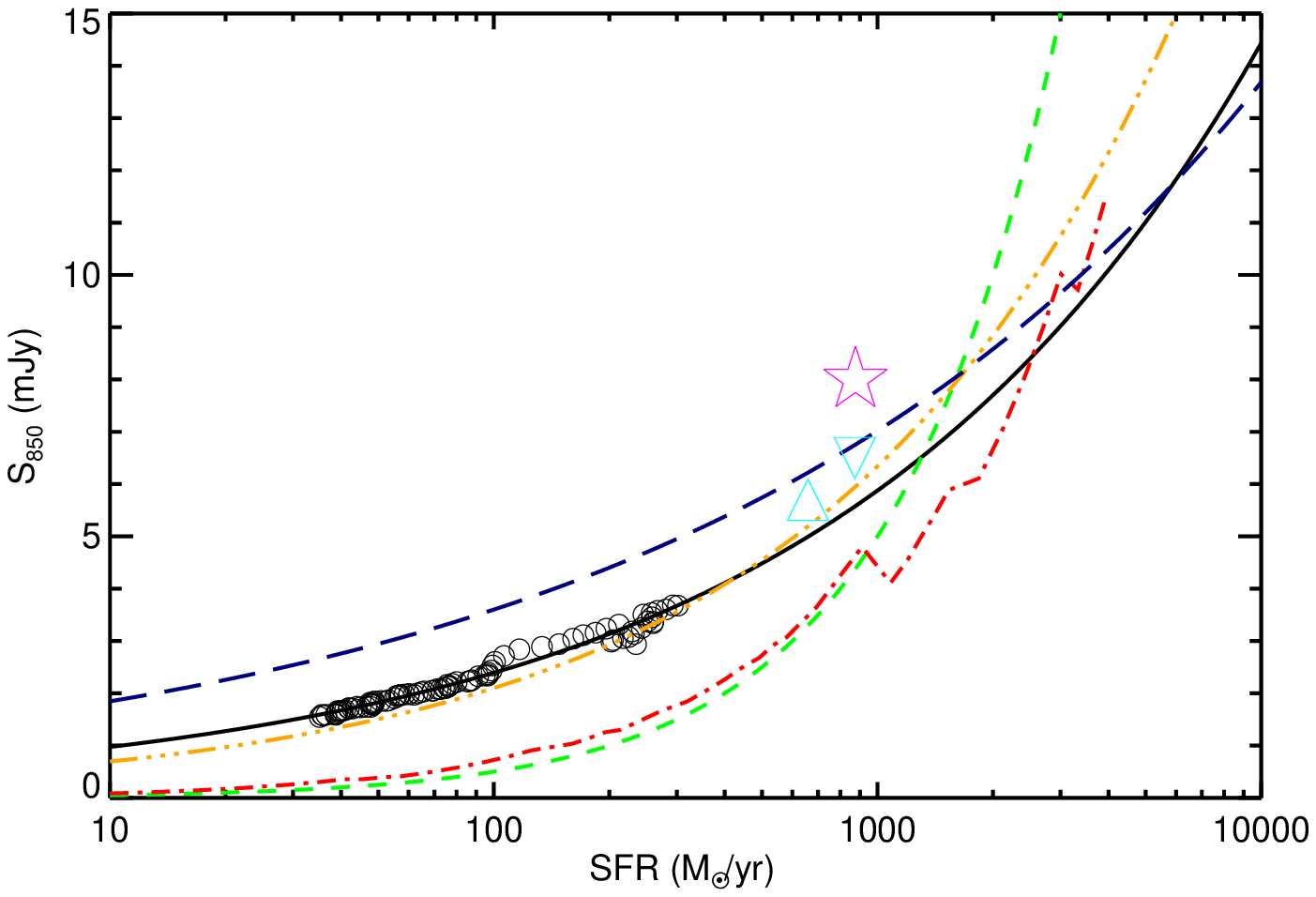}{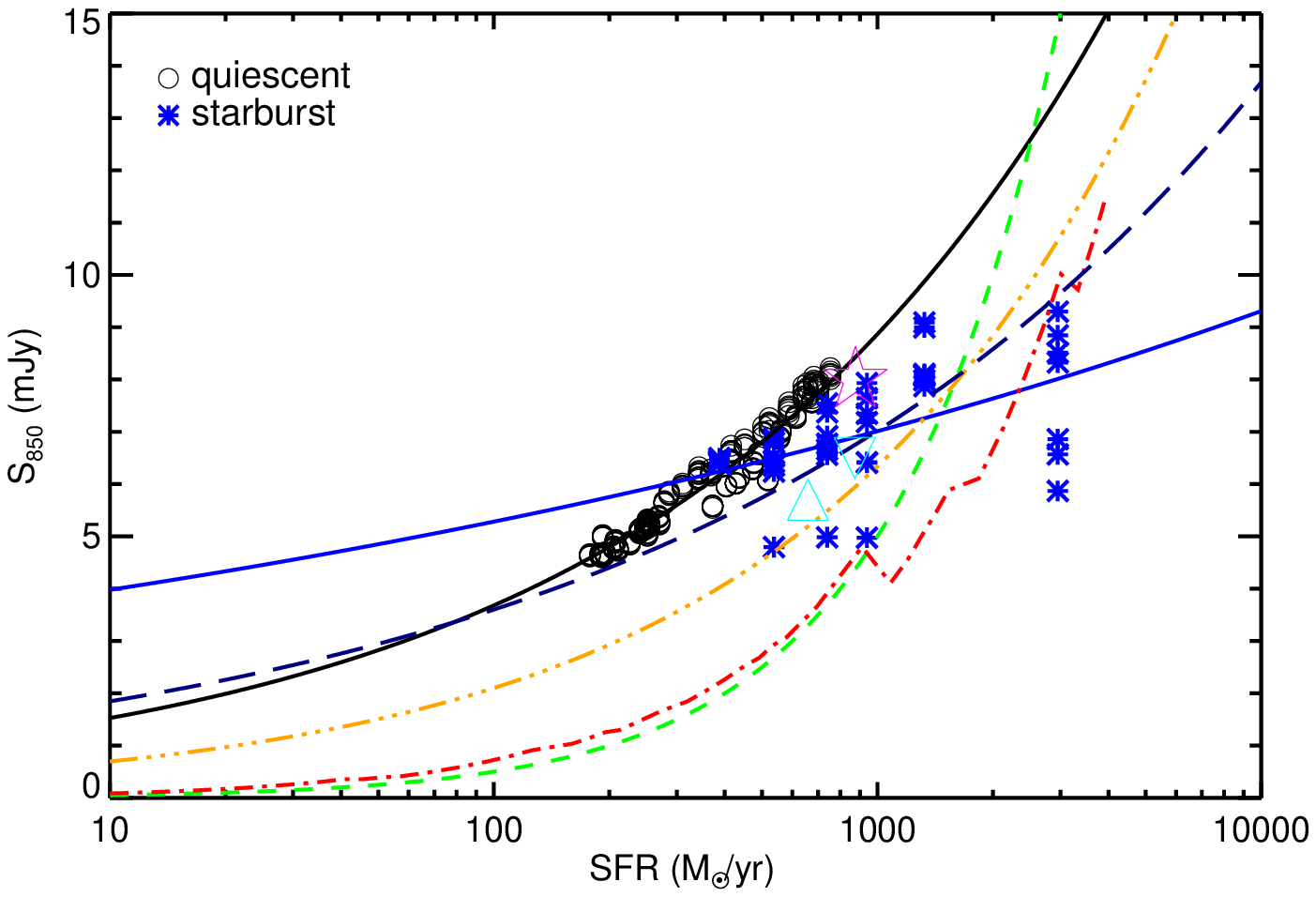}
\caption{Integrated SCUBA 850 \micron ~flux density (mJy) versus SFR ($\msunperyr$) for the isolated disk (\emph{left}) and major merger
(\emph{right}) viewed from all of the 7 different cameras.
The best-fit power laws (solid lines), linear relation $S_{850} = 0.5 {\rm ~mJy} ~({\rm SFR}/100 ~\msunperyr)$
(green dashed lines; normalization from \citealt{Neri:2003} and Equation \ref{eq:K98_IR_SFR_cal}),
the relation for the \citet{Chary:2001} templates (red dash dot), the \citet{Magnelli:2010} relations for their entire sample (orange dash dot dot dot) and excluding
the lensed SMGs (navy long-dashed), the value for the \citet{Pope:2008MIR} composite SED (magenta star),
and the median (cyan upward-pointing triangle) and mean (cyan downward-pointing triangle) from \citet{Michalowski:2010masses} are also shown.
For the isolated disk, the submm flux density is tightly correlated with both SFR
and $\lbol$, increasing essentially monotonically as SFR$^{0.4}$. For the major merger, pre-coalescence (black open circles) the power-law index is the
same as for the isolated disks because the SFR, $\lbol$, dust mass, and submm flux density are essentially two times the isolated disk values (see Figure
\ref{fig:iso_evolution}), so only the normalization changes. During the coalescence-induced starburst (blue asterisks), the relationship is significantly
shallower, with submm flux density scaling as SFR$^{0.1}$. This is due to two effects: 1. The stars formed before the peak of the starburst contribute significantly to $\lbol$
at the starburst peak, so $\lbol \not\propto$ SFR. 2. The rapid gas consumption during the burst causes $\mdust$ to plummet, and
the decrease in dust mass and more compact geometry cause $\tdust$ to increase
sharply, mitigating the increase of $S_{850}$ caused by the increased $\lbol$.}
\label{fig:850_vs_sfr}
\end{figure*}

Figure \ref{fig:850_vs_sfr} shows the observed SCUBA 850 \micron ~flux density in mJy versus star formation rate
in units of $\msunperyr$ for the isolated disk (\emph{left}) and major merger (\emph{right}) viewed from all of the 7 cameras.
The submm flux density of the isolated disk
is tightly correlated with SFR, increasing monotonically as SFR$^{0.4}$ (see best-fit curve). This correlation occurs because once the disk settles,
$\lbol$ and SFR both decrease exponentially with time. The dust mass also decreases, but by less than a factor of 2
over the 2 Gyr of the simulation (see Figure \ref{fig:iso_evolution}). Both the decreased luminosity and the decreased dust mass cause the
submm flux density to decrease. 

The case for the major merger is again qualitatively different. Pre-coalescence, the relationships are essentially the same as for the isolated disks.
This is because
$S_{850}$, SFR, $\lbol$, and $\mdust$ at this stage are essentially just the sum of the two disks' isolated values, and multiplying all quantities by the same
factor (2 for the major merger here) does not change the power-law index. The normalization of the relation is $\sim 1.5$ greater for the merging disks than
for the isolated disk. The reason is as follows: An isolated disk of SFR $s$ has $S_{850} = A s^{0.4}$, where $A$ is the normalization of the SFR-$S_{850}$ relation
for the isolated disk. For a non-interacting system of two identical disks to have total SFR equal to that of the single isolated disk, the two disks must each have
SFR = $0.5 s$. Thus the total submm flux density of the system is the submm flux density of a single disk of SFR $0.5 s$, which we calculate using the isolated relation, multiplied
by 2. This is $S_{850} = 2 A (0.5s)^{0.4} = 1.5 A s^{0.4}.$ Therefore the normalization of the SFR-$S_{850}$ relation for the sum of two identical disks is
1.5 times that of the individual disk relation. This fact has important implications for the SMG population, which we discuss in \S\ref{S:smg_bimodality}.

On the other hand, the merger-induced burst is significantly less effective at boosting the submm flux density. For a given SFR, the submm flux density is significantly
less than for the isolated and pre-coalescence (quiescent star formation) cases. This is because of two reasons: 1. The sharp decrease in dust mass
and more compact geometry cause an increase in dust temperature, mitigating the increase in $S_{850}$ caused by increased $\lbol$.
2. The significant luminosity contributed by stars formed before the starburst causes $\lbol$ to increase sub-linearly with SFR.\footnote{In principle the AGN can also
cause such an effect, but for snapshots classified as SMGs the typical AGN contribution to the IR luminosity is $\la 10$\%, so the AGN is sub-dominant.}
During the burst,
$\lbol \approx L_{\rm pre-peak} + \alpha {~\rm SFR}$, where $\alpha$ is the luminosity per unit star formation rate for an instantaneous burst. Thus $\lbol$
is not proportional to SFR when $L_{\rm pre-peak}$ is non-negligible compared to the luminosity of newly-formed stars, which is the case
here because a relatively small fraction of the stellar mass is formed in the sharp, short-duration burst.
For the burst, the submm flux density scales as SFR$^{0.1}$ (see best-fit curve), significantly more weakly than for quiescent star formation,
and the ratio of submm flux density to SFR is significantly lower.
Hence, bursts of star formation are significantly less effective at boosting submm flux density than one might naively expect.

It is interesting to note that, during the starburst, the observed submm flux density can vary significantly with viewing angle (e.g., for the
snapshot with peak SFR, $S_{850}$ varies in the range $\sim 6 - 9$ mJy depending on the camera). We have confirmed that this variation is due to
dust self-absorption: the central regions of the starburst can be so obscured that even the IR emission is significantly anisotropic.
As a result, the dust temperature, and thus submm flux density, depends on the line-of-sight. Though we will not explore this possibility further in
this work, we note that differences in viewing angle may be enough to account for the spread of dust temperatures observed for
high-z ULIRGs. In other words, from one viewing angle a simulated galaxy may be identified as an SMG whereas from another viewing
angle the same galaxy could be identified as a hot-dust-dominated ULIRG undetected in the submm.

Figure \ref{fig:850_vs_sfr} also shows the linear relation $S_{850} = 0.5 {\rm ~mJy} ~({\rm SFR}/100 ~\msunperyr)$
(green dashed lines; obtained using the $S_{850}-\lir$ relation from \citealt{Neri:2003} and Equation \ref{eq:K98_IR_SFR_cal}),
the relation for the \citet{Chary:2001} templates (red dash dot), the \citet{Magnelli:2010} relations for their entire sample (orange dash dot dot dot) and excluding
the lensed SMGs (navy long-dashed), the value for the \citet{Pope:2008MIR} composite SED (magenta star),
and the median (cyan upward-pointing triangle) and mean (cyan downward-pointing triangle) from \citet{Michalowski:2010masses}. 
The \citet{Chary:2001} and \citet{Pope:2008MIR} values were obtained by redshifting the templates to $z = 2$ and converting $\lir$ of each template to SFR
using Equation \ref{eq:K98_IR_SFR_cal}. We used Equation \ref{eq:K98_IR_SFR_cal} to convert the \citet{Magnelli:2010}
relations from $\lir$ to SFR.

The typical values
from \citet{Pope:2008MIR} and \citet{Michalowski:2010masses} are consistent with the data from our major merger simulation. As explained above,
the relations we find are much shallower than linear, so the \citet{Neri:2003} relation differs significantly from our relations for both the quiescent and
starburst modes. The \citet{Chary:2001} templates are also very discrepant because high-redshift ULIRG SEDs are often better
fit by local templates appropriate for less luminous, colder galaxies \citep[e.g.,][]{Pope:2006,Dannerbauer:2010,Rex:2010}.
The \citet{Magnelli:2010} relations agree better with our simulations: for the full sample, $S_{850} \propto$ SFR$^{0.48}$. This is a slightly steeper 
relation than what we find for quiescent disks and significantly steeper than that for starbursts. 
When the six lensed SMGs are removed from their sample, \citeauthor{Magnelli:2010} find $S_{850} \propto$ SFR$^{0.29}$. The power-law
index of this relation is less than that for our quiescent disks but greater than that for our starbursts. The lensed SMGs tend to be intrinsically
fainter and thus less likely to be strong starbursts than the non-lensed population, so it is reasonable that inclusion of the lensed SMGs
leads to a steeper relation. While it is interesting that the \citeauthor{Magnelli:2010} relation for the unlensed SMGS is crudely consistent with what we expect for a
mixed population, one should not over-interpret this comparison. As we will discuss below, $S_{850}$ cannot be determined solely from SFR because
the dust mass plays a significant role also. However, a robust conclusion that should be drawn from Figure \ref{fig:850_vs_sfr} is that, in the simulations, the starburst mode
is less efficient at boosting submm flux than the quiescent mode.

\subsection{Dependence of submm flux density on SFR, $\lbol$, and $\mdust$}

For a given SFR, galaxies of different masses tend to also have different dust masses;
thus the normalization of the $S_{850}$-SFR relation varies for different mass simulations though the scalings are similar.
As a result, one cannot calculate the submm flux density given only the SFR, but it is possible to parameterize the submm flux density as a function of SFR and
dust mass. Since much of the discrepancy in the $S_{850}$-SFR relations for quiescent star formation and starbursts is caused by the rapid decrease
in dust mass during the starburst, we expect that including dust mass in our parameterization will eliminate much of the difference between
quiescent and starburst star formation modes. We have analyzed the full set of simulations from our
SMG number counts work \citep[][C. Hayward et al.~2011, in preparation]{Hayward:2011num_cts_proc}, which includes a range of progenitor disk baryonic
masses ($\sim 3.5 \times 10^{10} - 4 \times 10^{11} M_{\odot}$), mass ratios ($\sim 0.1 - 1$), and initial gas fractions ($0.6 - 0.8$),
fitting the submm flux density as a power law in both SFR and dust mass. Both the quiescent and starburst phases are included.
Perhaps surprisingly, the following relations (for simulated galaxies placed at redshift $z = 2$)
hold to within $\sim0.1$ dex for all but a few outliers over the range $0.5 {\rm ~mJy} \la S_{850} \la 15$ mJy:
\begin{eqnarray} \label{eq:sfr_fitting_functions}
S_{850} &=& 0.65 {\rm ~mJy} \left(\frac{\rm ~SFR}{100 ~\msunperyr}\right)^{0.42} \left(\frac{\mdust}{10^8 \msun}\right)^{0.58} \\
S_{1.1} &=& 0.30 {\rm ~mJy} \left(\frac{\rm ~SFR}{100 ~\msunperyr}\right)^{0.36} \left(\frac{\mdust}{10^8 \msun}\right)^{0.61}, \nonumber
\end{eqnarray}
where $S_{850}$ and $S_{1.1}$ are the fluxes in the SCUBA 850 $\micron$ and AzTEC 1.1 mm bands, respectively.

Note that the SFR exponent
in these relations is similar to that for the $S_{850}$-SFR relation for quiescent SF ($\sim 0.4$). As explained in \S\ref{S:S850-SFR}, the sharp decrease in dust mass during the
starburst is one of the main reasons the $S_{850}$-SFR relation is much shallow for starbursts than for quiescent SF. Adding $\mdust$ as
a parameter effectively decouples this effect; in other words, \emph{for fixed dust mass} the $S_{850}$-SFR relations for the two SF modes are much more similar
than when the evolution of the dust mass is taken into account. Remaining differences caused by the contribution from stars formed pre-burst to the burst luminosity,
AGN contribution, geometry, and other factors prevent Eq. (\ref{eq:sfr_fitting_functions}) from recovering $S_{850}$ exactly, but the small scatter suggests that
these factors are subdominant.

\begin{figure*}
\epsscale{1.15}
\plottwo{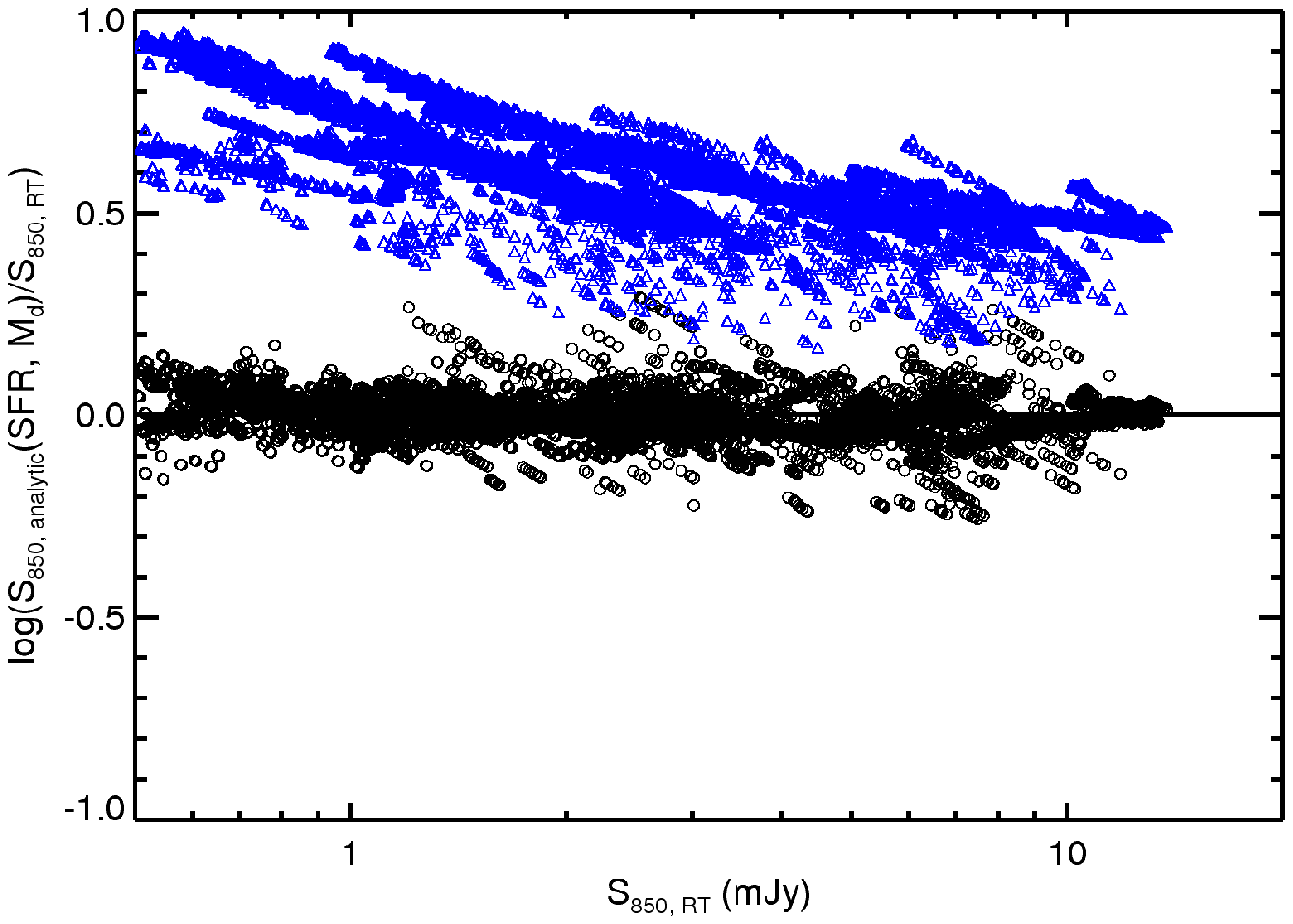}{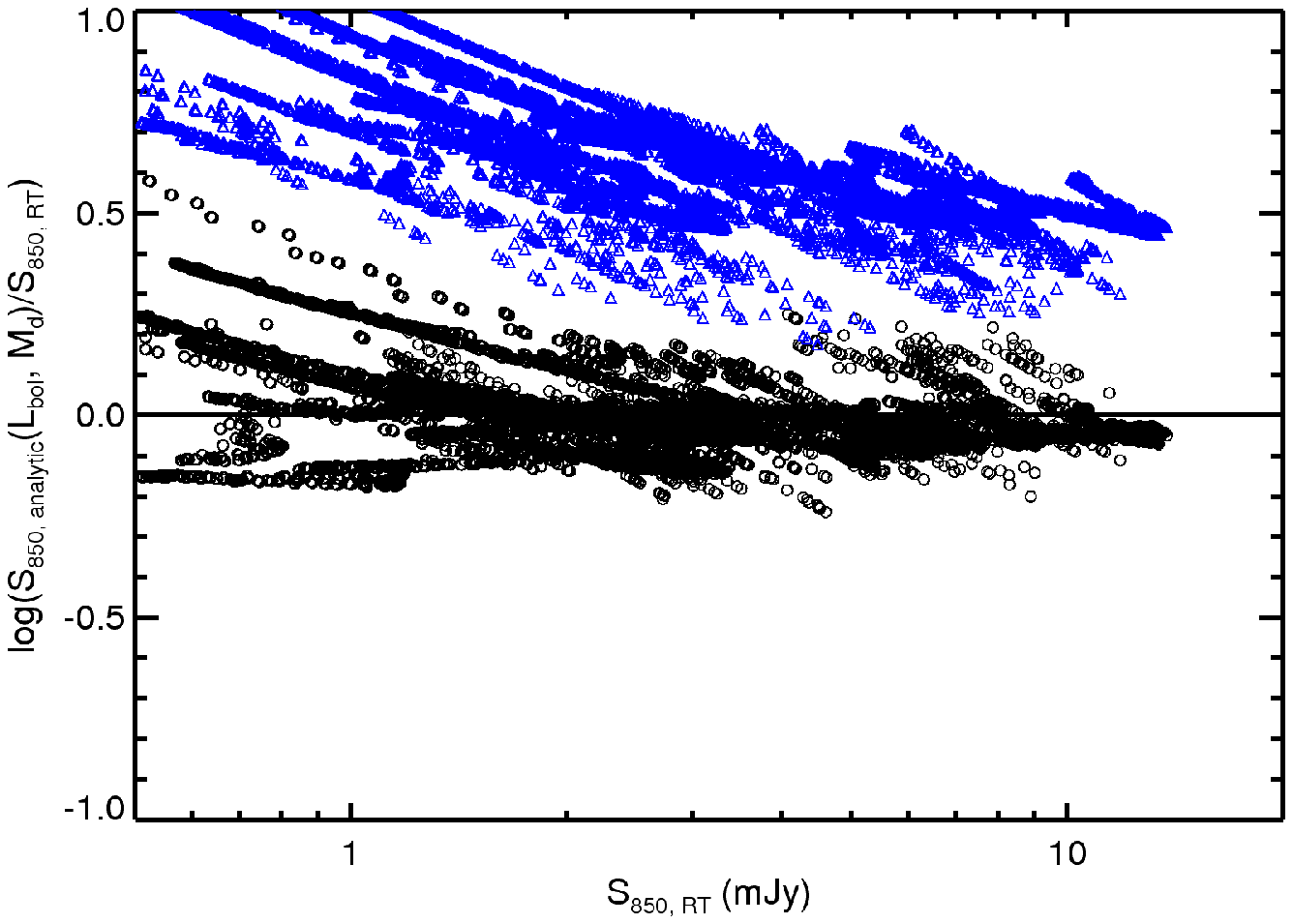}
\caption{\textit{Left:} Logarithm of the ratio of the submm flux density calculated using one of the analytic forms ($S_{\rm 850, analytic})$ to the submm flux density
calculated through the full RT $(S_{\rm 850, RT})$ vs. $S_{\rm 850, RT}$ for all time snapshots of our simulated galaxies.
Black circles show the ratio when $S_{\rm 850, analytic}$ is calculated from the SFR and $\mdust$ of our simulated galaxies
using Equation (\ref{eq:sfr_fitting_functions}). Blue triangles show the ratio when the optically thin modified blackbody model (Equation \ref{eq:single_T_SFR_scaling}) is used.
The line $S_{850,\rm analytic} = S_{\rm 850, RT}$ is shown to guide the eye.
\textit{Right:} Same, except now the black circles show the values when Equation (\ref{eq:L_fitting_functions}) is used to
calculate the submm flux density from $\lbol$ and $\mdust$ of the simulation snapshots, and the blue triangles show the values when Equation (\ref{eq:single_T_L_scaling}) is used,
assuming $\lbol \approx \ldust$. In both cases the simple optically thin modified blackbody model overpredicts the submm flux density by $\ga0.3$ dex
for the simulated SMGs, and the overprediction is worse for lower $S_{850}$.}
\label{fig:fitting_functions}
\end{figure*}

We can also fit the submm flux density as a function of $\lbol$ and $M_d$:
\begin{eqnarray} \label{eq:L_fitting_functions}
S_{850} &=& 0.40 {\rm ~mJy} \left(\frac{\lbol}{10^{12} \lsun}\right)^{0.52} \left(\frac{\mdust}{10^8 \msun}\right)^{0.60} \\
S_{1.1} &=& 0.20 {\rm ~mJy} \left(\frac{\lbol}{10^{12} \lsun}\right)^{0.46} \left(\frac{\mdust}{10^8 \msun}\right)^{0.63}. \nonumber
\end{eqnarray}
These fitting functions are accurate to within $\sim 0.15$ dex. Replacing $\lbol$ with the bolometric IR luminosity $\lir$
yields a similar result.
Figure \ref{fig:fitting_functions} shows how well these fitting functions reproduce the submm flux density of our simulated galaxies. It is discussed in
more detail in \S\ref{S:comparison_of_relations}.

\subsubsection{Relations for an optically thin modified blackbody} \label{S:mod_BB_relations}

It is instructive to compare the above relations to those for a single-temperature mass of dust transparent to its own emission, the model
which is implicit in the standard method of fitting a modified blackbody to the IR SED.
For a mass of dust in thermal equilibrium with temperature $\tdust$ we can express the submm flux density as a function of dust bolometric luminosity $\ldust$
and dust mass $\mdust$ or SFR and $\mdust$. Assuming $z = 2$ and far-IR spectral index $\beta = 2$, the relations are (see the Appendix for a derivation)
\begin{equation} \label{eq:single_T_SFR_scaling}
S_{850} = 1.4 {\rm ~mJy} \left(\frac{\rm ~SFR}{100 ~\msunperyr}\right)^{1/6} \left(\frac{\mdust}{10^8 \msun}\right)^{5/6}
\end{equation}
and
\begin{equation} \label{eq:single_T_L_scaling}
S_{850} = 1.4 {\rm ~mJy} \left(\frac{\ldust}{10^{12} \lsun}\right)^{1/6} \left(\frac{\mdust}{10^8 \msun}\right)^{5/6}.
\end{equation}
These should be compared to Equations (\ref{eq:sfr_fitting_functions}) and (\ref{eq:L_fitting_functions}), respectively, by assuming $\lbol \approx \ldust$ (i.e.,
the luminosity emitted by stars and AGN is completely absorbed by dust), which is a reasonable approximation for snapshots classified as SMGs ($S_{850} > 3$ mJy).
The submm flux depends only weakly on redshift for the redshift range of interest $(1 \la z \la 5)$, scaling as $(1+z)^{\beta-1} D_A^{-2}$, where $D_A$ is the angular
diameter distance at redshift $z$ (see Equation \ref{eq:S_nu_general}).

\subsubsection{Comparison of the relations to the full radiative transfer} \label{S:comparison_of_relations}

Figure \ref{fig:fitting_functions} shows the logarithm of the ratio of the submm flux density calculated using the above equations ($S_{\rm 850, analytic}$) to the submm flux density
calculated through the RT ($S_{\rm 850, RT}$) versus $S_{850, RT}$. The left panel shows the results when Equations (\ref{eq:sfr_fitting_functions}, black circles) and
(\ref{eq:single_T_SFR_scaling}, blue triangles) are used to calculate $S_{850}$ from SFR and $\mdust$. The right panel shows the results when $\lbol$ and $\mdust$ are
used instead, with the black circles corresponding to Equation (\ref{eq:L_fitting_functions}) and the blue triangles Equation (\ref{eq:single_T_L_scaling}). 
The fitting functions derived from the simulations are able to reproduce $S_{850}$ from SFR ($\lbol$) and $\mdust$ to within $\sim 0.1 ~(0.15)$ dex.
The simple modified blackbody model tends to overpredict the submm flux density by $\ga 0.3$ dex; the typical over-prediction is
$\sim 0.5$ dex (a factor of 3).
(Note that the corresponding uncertainties in SFR, $\lbol$, and $\mdust$ for fixed observed submm flux density would be less because the submm flux density scales with
these quantities sublinearly.)
Furthermore, the error in the prediction correlates with SFR ($\lbol$) and dust mass because of the differences in the power-law indices for the fitting functions
and the modified blackbody relations.

The optically thin modified blackbody model fails for multiple reasons. From SED fitting we find that the simulated galaxies can have effective optical depth $\tau > 1$
out to rest-frame $\sim 200$ \micron. For fixed dust temperature,
the optically thin assumption will result in an overestimate of the luminosity density at wavelengths for which $\tau \ga 1$  because $(1 - \exp[-(\nu/\nu_0)^{\beta}])
< (\nu/\nu_0)^{\beta}$ for all $\nu > 0$. Thus Equation (\ref{eq:L_dust}) will overestimate $L_d$ for fixed $T_d$ and $M_d$. If $L_d$ and $M_d$ are fixed, the dust
temperature will be \textit{underestimated} when optical thinness is assumed, and, therefore, the submm flux density will be overestimated.
Also, $\ldust \approx \lbol$ is less accurate an approximation for the faintest sources than for the brightest because $L_d/\lbol$ increases with $\lbol$ \citep{Jonsson:2006}.
If $\ldust < \lbol$, the assumption that $\ldust \approx \lbol$
will overestimate $L_d$ and thus overestimate the submm flux density; this may explain why the overprediction is worse for lower $S_{850}$.
Finally, we have seen above that the assumption $\lbol \propto$ SFR is invalid during the burst.

\section{Discussion} \label{S:discussion}

We have demonstrated that the submm flux density of a galaxy scales differently with SFR for quiescent star formation and starbursts. The ratio of
sub-mm flux density to SFR is significantly less for merger-induced bursts than for quiescent star formation. This is because of the rapid decrease in dust
mass and more compact geometry during the starburst, which causes the SED to become hotter, and the significant contribution from stars formed pre-burst to the luminosity
during the burst, which makes the luminosity
increase by a much smaller factor than the SFR. As a result, merger-induced starbursts are less efficient at boosting submm flux density than one
might naively expect.

Our results have a number of important implications; we discuss these now.

\subsection{Predicting submm flux densities from models}

One implication of this study is that, at a fixed redshift, the galaxies with highest submm flux density are not necessarily those with the highest bolometric luminosities
or SFRs. Thus theoretical models, be they simulations or semi-analytic, must
explicitly calculate the submm flux density of their simulated galaxies in order to select which are SMGs
as opposed to simply selecting the most rapidly star-forming
or most luminous objects. However, the computational expense required to self-consistently calculate the submm flux density
limits this approach to idealized, non-cosmological simulations (as done here), individual galaxies excised from cosmological
zoom-in simulations, or semi-analytic models in which various simplifying assumptions
must be made. As an alternative to performing RT, cosmological simulations and semi-analytic models
can use the relations among submm flux density, SFR or bolometric luminosity, and dust mass presented herein (Equations \ref{eq:sfr_fitting_functions}
and \ref{eq:L_fitting_functions}) to assign submm flux density to simulated galaxies. Additionally, observers can use the relations
to estimate the instantaneous SFR given submm flux density and dust mass (obtained from fitting the IR SED using the full modified blackbody as we
have done or by measuring the gas mass and assuming a dust-to-gas ratio; ignoring uncertainties on submm flux density,
a dust mass accurate to a factor of 2 gives SFR accurate to a factor of 3). 

\subsection{Heterogeneity of the SMG population} \label{S:smg_bimodality}

These results also imply that the SMG population is heterogeneous. We have seen that it is possible for a quiescently star-forming
disk to have submm flux density equal to that of a merger with much higher SFR (Figure \ref{fig:850_vs_sfr}).
Furthermore, since the scaling of submm flux density with SFR is sublinear, adding two equal disks (and thus doubling the
dust mass and SFR of the system) increases the submm flux density more than simply boosting the SFR by a factor of 2;
Equation (\ref{eq:sfr_fitting_functions}) shows that SFR would have to be boosted by 5x to achieve a 2x boost in submm flux density
if dust mass is kept constant.
When the sharp increase in dust temperature during the starburst and the narrowness of the burst are accounted for the effect becomes even
stronger: in Figure \ref{fig:iso_evolution} we see that a $\ga 16$x increase in SFR gives a $\la 2$x increase in submm flux density.

The single-dish submm telescopes used for wide-field surveys of SMGs have beam sizes $\ga 15'' (\ga 130$ kpc at $ z = 2$). Thus, during
a merger the two progenitor galaxies will spend a considerable amount of time within the area of the beam. From the above arguments,
we see that this is a very efficient way to create an SMG, but this contribution has been relatively unappreciated.
We argue that the SMG population attributable to mergers is bimodal: some are merger-induced starbursts and some are two (or more) infalling disks
(normal galaxies that are not yet interacting strongly) blended into one submm source (``galaxy-pair SMGs'').\footnote{\citet{Wang:2011} recently
presented high-resolution submm continuum images of two SMGs which were previously identified as single sources but are resolved
as 2 or 3 distinct submm sources in their images. The sources are at significantly different redshifts and thus physically unrelated.
Our galaxy-pair SMGs are also two sources blended into one, but they are distinct from the type of SMGs Wang et al.~observed because
they are merging and thus physically connected. Both types of blended sources are potentially important SMG subpopulations
that complicate our understanding of SMGs, and it is crucial to understand the relative contributions of merger-induced starbursts,
galaxy pairs/infall-stage mergers, quiescently star-forming disks, and physically unrelated, blended sources.} Note also that not only major mergers but
also favorably oriented minor mergers \citep[see, e.g.,][]{Cox:2008,Hopkins:2009disk_survival} can contribute to the SMG population.

Furthermore, the most massive, rarest `isolated disks', and even small groups, may also contribute; we expect this contribution to be
subdominant because SMGs are on the exponential tail of the mass function, but we defer a precise determination of their contribution to future work.
Though both the merger-induced starburst and galaxy-pair populations are mergers, only in the former is the star formation merger-driven (and only partially, as
the baseline star formation that would occur in the disks even if they were not interacting is significant). Given that the physically meaningful property of local
ULIRGs is not that their IR luminosities are $\ge 10^{12} \lsun$ but that they are powered by merger-driven star formation and AGN, only the
merger-driven starburst category of SMGs should be considered physically analogous to local ULIRGs.

The observational signatures and physical implications of this bimodality will be discussed in future work. Here we simply note that the galaxy-pair
contribution is supported observationally by the frequency of multiple radio \citep{Ivison:2002,Ivison:2007,Chapman:2005,Younger:2009SMG_interf},
24 \micron~ \citep{Pope:2006}, and 350 \micron~ \citep{Kovacs:2010}
counterparts to SMGs; by CO interferometry showing that a large fraction of SMGs are resolved binaries \citep{Tacconi:2006,Tacconi:2008,Bothwell:2010,Engel:2010};
and by the SMGs that have morphologies that do not resemble merger remnants \citep[e.g.,][]{Bothwell:2010,Carilli:2010,Ricciardelli:2010,
Targett:2011}.\footnote{Note, however, that
the \citet{Bothwell:2010} and \citet{Carilli:2010} objects that resemble disk galaxies may in fact be the molecular disks that re-form rapidly after a
gas-rich merger \citep{Narayanan:2008quasar_CO,Robertson:2008}.}

\subsection{SMG masses}

The masses of SMGs are hotly debated, with different authors finding masses discrepant by $\sim 6$x for the same SMGs \citep{Michalowski:2010masses,
Michalowski:2010production,Hainline:2010}. Accurate masses are important in order to test potential evolutionary relationships among SMGs and other
galaxy classes \citep[e.g.,][]{Brodwin:2008,Bussmann:2009morphology,Bussmann:2009dog,Narayanan:2010dog,Rothberg:2010}
and to check that number densities of SMGs are consistent with observed stellar mass functions. Stellar mass determinations from SED fitting
are limited by uncertainties in stellar evolution tracks, the initial mass function, star formation histories, dust attenuation, and AGN contamination. Since
our models use star formation histories, attenuation (from the geometry of stars, AGN, and dust), and AGN components that originate
directly from the hydrodynamic simulations instead of the standard assumptions---e.g., instantaneous burst or exponential star formation histories,
the Calzetti attenuation law \citep{Calzetti:1994,Calzetti:2000,Calzetti:1997}---we can provide constraints on SMG stellar masses
that are complementary to those derived from SED fitting.

Given the inefficiency of bursts at boosting submm flux density that we have demonstrated above, SMGs must be very massive, because smaller
galaxies undergoing even very strong bursts cannot make SMGs. Our models require $M_{\star} \ga 6 \times 10^{10} \msun$ to reach
$S_{850} \ga 3$ mJy, and typical masses are higher. The area of the $S_{850}-M_{\star}$ plot spanned by our models agrees well with the
observationally derived values of \citet{Michalowski:2010masses}. About half of the \citet{Hainline:2010} values lie in the area spanned by our models,
whereas the other half have lower masses. However, the single-component star formation histories assumed by Hainline et al.~may cause the
stellar masses to be underestimated by $\sim 2$x, which would resolve much of the discrepancy. A detailed comparison of the mass estimates
will be presented in M. Micha{\l}owski et al. (2011, submitted).

\subsection{SMG duty cycles}

Understanding the duty cycle of SMGs is important for predicting submm counts from models, quantifying the contribution of SMGs to
stellar mass buildup, and interpreting star formation efficiencies of SMGs. Since the submm flux density depends on luminosity heating the dust, dust mass,
and geometry, the submm duty cycle depends on the same factors. As we have seen, the starburst induced at merger coalescence causes a
sharp peak in SFR, $\lbol$, and submm flux density. However, the duty cycle is limited because of the sharp cutoff in SFR, and thus drop in $\lbol$, after
the burst and the significant drop in dust mass that occurs as highly enriched gas is consumed in the burst.

Since the submm flux density depends more strongly on dust mass than either SFR or $\lbol$ (see Equations \ref{eq:sfr_fitting_functions}
and \ref{eq:L_fitting_functions}), it is more efficient if one can keep more dust around at the expense of lower star formation rate. The
quiescent star formation mode does exactly this. As a result, the galaxy-pair phase (discussed in \S\ref{S:smg_bimodality}) adds significantly to
the submm duty cycle. Figure \ref{fig:iso_evolution} shows that the galaxy-pair phase has a longer submm duty cycle than the burst,
though the lack of smooth accretion in our simulations---and thus need for starting with very high initial gas fractions---complicates a precise determination
of the relative flux densities and duty cycles of the galaxy-pair and starburst phases. Regardless, it is clear that the galaxy-pair phase increases the SMG
duty cycle significantly, alleviating some of the tension between the submm counts estimated from high-redshift major merger rates and short
($\sim 100$ Myr) starburst duty cycles by \citet{Dave:2010} and the observed submm counts.
Inclusion of both SMG populations is crucial to match the observed SMG number counts without resorting to a top-heavy
initial mass function \citep[][C. Hayward et al.~2011, in preparation]{Hayward:2011num_cts_proc}.

\subsection{Implications for IR SED fitting}

One reason the single-temperature, optically thin modified blackbody fails is that the effective optical depth of our simulated SMGs can be greater than 1 out to
rest-frame $\sim 200~\micron$. This is consistent with the effective optical depths derived by \citet{Lupu:2010} and \citet{Conley:2011} when they fit a general modified blackbody (i.e., not
assuming optical thinness) to the IR SEDs of their SMGs. By assuming optical thinness and only fitting longward of the FIR peak, one overestimates the luminosity
density at wavelengths for which $\tau \ga 1$ (see \S\ref{S:comparison_of_relations}).\footnote{This may explain
why \citet{Pope:2006} found that the submm flux density tends to overpredict $\lir$, and the overprediction is worse for SMGs at low redshift where the SED is sampled further
from the IR peak.} Consequently, the assumption of optical thinness yields a colder dust temperature (by as much as $\sim 20$ K) than if optical thinness is assumed.

In the pre-\textit{Herschel} era, often flux densities in only a few FIR bands and at 1.4 GHz were available for large samples of SMGs. As a result of the limited number of
data points, models more complex than the optically thin modified blackbody (e.g., the full modified blackbody used here or models assuming a distribution of
temperatures) could not provide a better description of the IR SEDs \citep[e.g.,][]{Kovacs:2006}. Recently, some authors have used models incorporating a
distribution of temperatures \citep[e.g.,][]{Kovacs:2010,Michalowski:2010masses}, finding that such models better described the IR SEDs \citep{Kovacs:2010}.
Now that \textit{Herschel} PACS \citep{Dannerbauer:2010,Magnelli:2010}
and SPIRE \citep{Chapman:2010} have provided data over the entire FIR SED for large samples of SMGs it is possible to perform more sophisticated fitting
for many SMGs.
Given the physical inferences that are drawn from effective dust temperatures obtained via FIR SED fitting, it is important to have as robust a method as possible
and to take full advantage of the available data. We will present such a method in future work.

\subsection{Limitations of our model}

At this point we find it instructive to define the limitations of this work so that the results can be placed in an appropriate context
and future experiments can be designed for maximal impact.  One of the primary limitations, both for the radiative transfer
and the hydrodynamics, is the treatment of the sub-resolution interstellar
medium, especially---because we focus upon the submm flux---the structure, distribution, and composition of the dust which
dominates emission at submm wavelengths. In fact, the differences between the model employed in this work and that we used in
N10 were motivated by a desire for a simpler treatment of the sub-resolution ISM (see \S\ref{S:diff_from_nar}).
In N10, stars with age $\le 10$ Myr dominated the submm flux, so the submm flux was closely tied to the SFR (see Fig. 1 of N10).
Our simplified assumptions (no sub-resolution PDR model, uniform ISM density on scales below the SPH smoothing length) result in submm flux
that is tied more directly to the bolometric luminosity than the instantaneous star formation rate because stars with age $>10$ Myr
contribute significantly to the bolometric luminosity and, because they are still deeply embedded in dust, the submm flux.

While our model has the advantages of simplicity and strict physical consistency (because the obscuration originates purely from the hydrodynamic
simulations rather than from sub-resolution PDRs), we still must make an assumption about the sub-resolution structure of the ISM.
By assuming uniform density on scales below the smoothing length we only include clumpiness that arises from the hydrodynamic
simulations, so this assumption may be considered conservative. The assumption is also simplistic, of course, because the real ISM has
significant structure on scales $\la 100$ pc. However, proper treatment of radiative transfer through a clumpy medium is a significant
area of research in and of itself \citep[e.g.,][]{Hobson:1993,Witt:1996,Varosi:1999a} and thus beyond the scope of this work.
A study of the effects of sub-resolution dust clumpiness on galaxy SEDs and efforts to devise a better treatment of sub-resolution
dust clumpiness in \sunrise are underway.

We also caution that modifications to the simple star formation prescription and ISM treatment in the hydrodynamical simulations themselves
could change the amplitude and duration of starbursts
\citep{Cox:2006feedback,Springel:2005feedback}. However, while changing the SF prescription or ISM treatment could change the relative contribution
of quiescent and starburst star formation modes to the star formation history of a given merger, this alone should not change the differences between quiescent star formation and
starbursts which lead to the significantly different relationships between SFR and submm flux. Changing the SF or ISM prescription could substantially alter the spatial
distribution of dust and stars and thereby modify the detailed relations between SFR/$\lbol$, $\mdust$, and $S_{850}$.
However, because geometry is relatively unimportant in setting the relations,
this uncertainty should have a relatively limited influence on our results. Furthermore, changing the feedback implementation can alter the evolution of the IR
SEDs \citep{Chakrabarti:2007}.
Ongoing and future studies, with
much higher resolution and more advanced tracking of the clumpy ISM \citep[e.g.,][]{Hopkins:2011self-regulated_SF}, will improve the predictive power of our models.

Furthermore, we stress that the simulations presented here are not cosmological. We adopt this approach
because it enables us to achieve the high resolution necessary to perform radiative transfer in order to accurately calculate
the submm flux density; to survey the parameter space of progenitor masses, mass ratios, and orbits;
and to avoid uncertainties in modeling realistic galaxy populations in a cosmological environment.
The primary drawback of this approach for our present purposes is the lack of gas
accretion, which cosmological hydrodynamic
simulations show is a significant driver of star formation for the high-redshift, massive galaxies with which we are concerned
\citep{Keres:2005,Keres:2009a,Dekel:2009nature}. Gas accretion can continually replenish the gas in the galaxy, maintaining relatively
high gas fractions and relatively constant star formation histories \citep[e.g.,][]{Dave:2010}.

Inclusion of cosmological gas accretion would alter the time evolution presented in Figures
\ref{fig:iso_evolution}, but it would not significantly alter the
differences between the quiescent and merger-induced burst modes of star formation. This is because mergers would still superimpose
a strong burst of star formation and sharp decrease in gas mass over the baseline evolution.
Furthermore, unless smooth accretion significantly affects the geometry of stars and dust in the simulated galaxies, it will not have
a significant effect on the relationship between submm flux, $\lbol$, and dust mass (the relationship between submm flux, SFR, and dust mass
may be more affected because the relation between SFR and $\lbol$ may be changed significantly).
Thus inclusion of cosmological gas accretion should not qualitatively alter our results.

\subsection{Connections to previous work}

In previous work, we developed a model relating the evolution of galaxies,
starbursts, and quasars
\citep{Hopkins:2006unified_model,Hopkins:2006quasar_LF,Hopkins:2008red_Es,Hopkins:2009bulge_demo,Somerville:2008}.
A principal conclusion from these analyses is that
while starbursts driven by gas-rich mergers can account for many
instances of unusual activity in galaxies, they provide only a minor
contribution to the star formation history of the Universe
\citep{Hopkins:2006LFs_SFH}.  Indeed, \citet{Hopkins:2010IR_LF} emphasize that
much of the star formation during galaxy interactions occurs in the
``quiescent'' mode and should not be counted as part of a merger-induced
starburst.  This is supported by the decomposition of the
light profiles of nearby ongoing mergers \citep{Hopkins:2008extra_light}
and local cusp \citep{Hopkins:2009cusps} and core \citep{Hopkins:2009cores}
ellipticals.  All of these objects exhibit evidence of ``excess'' central
light \citep{Rothberg:2004,Kormendy:2009}, indicative of relic starbursts
\citep{Mihos:1994,Mihos:1996} caused by merger-driven inflows of gas
\citep{Barnes:1991,Barnes:1996}.  The integrated mass in these components
agrees well with estimates of the cosmic history of merger-induced
starbursts \citep{HH:2010}.

The results presented herein extend these conclusions to high-redshift
phenomena.  Critically, we find that ``quiescent'' star formation
during galaxy interactions, i.e., star formation which occurs during the infall/pair stage,
is a key element in understanding the
brightest submm sources, especially their number counts and duty cycles,
and connecting them to other high-redshift populations including quasars
\citep{Hopkins:2008red_Es,Narayanan:2008quasar_CO}
and compact spheroidal galaxies \citep{Wuyts:2009b,Wuyts:2010}.
Just as nearby LIRGs represent a heterogeneous collection of merging
and isolated systems, it is natural to suggest that the population of
high-redshift submm galaxies is heterogeneous, as we have argued here.

\section{Conclusions}

We have combined high-resolution 3-D hydrodynamic simulations of high-redshift isolated and merging disk galaxies and 3-D Monte Carlo
dust radiative transfer calculations to study the submillimeter galaxy selection, focusing on the relationships among submm flux density, star formation rate,
bolometric luminosity, and dust mass. Our main conclusions are the following:

\begin{enumerate}
\item The relationship between SFR and submm flux density differs significantly for quiescent and starburst star formation modes. Starbursts produce
significantly less submm flux density for a given SFR, and the scaling between submm flux density and SFR is significantly weaker for bursts than for
quiescent star formation. Bursts are a very inefficient way to boost submm flux density (e.g., a starburst that increases SFR by $\ga 16$x increases submm flux density
by $\la 2$x). Another consequence is that the galaxies with highest submm flux density are not necessarily those with highest SFR or bolometric or infrared luminosity.

\item The submm flux density of our simulations can be parameterized as a power law in SFR and dust mass ($\lbol$ and dust mass) to within $\sim 0.1 (0.15)$
dex. The scaling derived from the commonly used optically thin modified blackbody model systematically overpredicts the submm flux density by $\ga 2$x
because numerous assumptions of the model (optical thinness in the FIR, $\lir \propto$ SFR, $\lbol \approx \lir$) do not hold.
The fitting functions we provide (Equations \ref{eq:sfr_fitting_functions} and \ref{eq:L_fitting_functions}) should be useful for calculating the flux density
in semi-analytical models and cosmological simulations when full radiative transfer cannot be performed and for interpreting observations.

\item Mergers create SMGs through another mechanism besides the strong starburst induced at coalescence---they cause the two infalling disks
to be observed as one submm source because both disks will be within the large ($\sim 15$'', or 130 kpc at $z = 2$) beam of the single-dish submm
telescopes used to identify SMGs during much of the infall stage. For major mergers, this effect boosts the submm flux density by 2x. To achieve the
same boost in submm flux density one would have to boost the SFR of a quiescent disk by $\sim 6$x or induce a starburst that boosts the SFR
by $\ga 16$x. This implies that the SMG population is heterogeneous: it is composed of both late-stage major mergers and two (or more)
infalling disks observed as a single submm source (``galaxy-pair SMGs''). The largest quiescently star-forming galaxies may also contribute.
Thus, unlike local ULIRGs, SMGs are a mix of quiescent and starburst sources.

\item SMGs must be very massive: to reach $S_{850} \ga 3$ mJy, stellar mass of at least $6 \times 10^{10} \msun$ is required, and typical
values are higher.

\item The submm duty cycles of our simulated galaxies are a factor of a few longer than what one would expect if all SMGs were merger-driven bursts
because the relatively gentle decline in SFR, $\lbol$, and dust mass during the galaxy-pair phase results in a longer duty cycle for the galaxy-pair
phase than for the starburst. The duty cycle of the latter is limited because the peak in luminosity is narrow and the dust temperature increases sharply
during the burst.

\item Fitting the SEDs of SMGs with an optically thin modified blackbody tends to yield significantly lower dust temperatures than when the full
opacity term is used because the effective optical depths can be $\sim 1$ out to rest-frame $\sim 200 ~\micron$, both for our simulated
SMGs and observed SMGs. Therefore, one should be cautious when interpreting effective dust temperatures derived via fitting an optically
thin modified blackbody to the FIR SED, especially when comparing SMGs to galaxies for which optical thinness in the IR may be a reasonable approximation.

\end{enumerate}

Future work will include predictions of submm number counts from our model, an investigation of the observational signatures and physical
implications of the proposed SMG bimodality, and an improved method for fitting IR SEDs of galaxies.

\acknowledgments

CCH thanks Scott Chapman, Emanuele Daddi, Helmut Dannerbauer, Phil Hopkins, Brandon Kelly, Chris McKee, Micha{\l} Micha{\l}owski,
Diego Munoz, Alex Pope, Barry Rothberg, Greg Snyder, and Josh Younger
for useful discussion and Jeff Pettibone for hospitality which facilitated the writing of this paper.
We thank Volker Springel for providing the non-public version of \gadgettwo used for this work and Brant Robertson for use of his code to generate
initial disk galaxies scaled to high redshift.
DK is supported by NASA through Hubble Fellowship grant HST-HF-51276.01-A. PJ acknowledges support by a grant from the W. M. Keck Foundation.
DN acknowledges support from a National Science Foundation Grant (AST-1009452).
The simulations in this paper were performed on the Odyssey cluster supported by the FAS Research Computing Group at Harvard University.

\bibliography{std_citations,smg}

\begin{thebibliography}{190}
\expandafter\ifx\csname natexlab\endcsname\relax\def\natexlab#1{#1}\fi

\bibitem[{{Alexander} {et~al.}(2005{\natexlab{a}}){Alexander}, Bauer, Chapman,
  Smail, Blain, Brandt, \& Ivison}]{Alexander:2005}
{Alexander}, D.~M., Bauer, F.~E., Chapman, S.~C., Smail, I., Blain, A.~W.,
  Brandt, W.~N., \& Ivison, R.~J. 2005{\natexlab{a}}, \apj, 632, 736

\bibitem[{{Alexander} {et~al.}(2005{\natexlab{b}}){Alexander}, {Smail},
  {Bauer}, {Chapman}, {Blain}, {Brandt}, \& {Ivison}}]{Alexander:2005b}
{Alexander}, D.~M., {Smail}, I., {Bauer}, F.~E., {Chapman}, S.~C., {Blain},
  A.~W., {Brandt}, W.~N., \& {Ivison}, R.~J. 2005{\natexlab{b}}, \nat, 434, 738

\bibitem[{{Alexander} {et~al.}(2008)}]{Alexander:2008}
{Alexander}, D.~M., {et~al.} 2008, \aj, 135, 1968

\bibitem[{Barger {et~al.}(1998)Barger, Cowie, Sanders, Fulton, Taniguchi, Sato,
  Kawara, \& Okuda}]{Barger:1998}
Barger, A.~J., Cowie, L.~L., Sanders, D.~B., Fulton, E., Taniguchi, Y., Sato,
  Y., Kawara, K., \& Okuda, H. 1998, \nat, 394, 248

\bibitem[{Barnes \& Hernquist(1991)}]{Barnes:1991}
Barnes, J., \& Hernquist, L. 1991, \apjl, 370, L65

\bibitem[{Barnes \& Hernquist(1996)}]{Barnes:1996}
---. 1996, \apj, 471, 115

\bibitem[{Barnes \& {Hut}(1986)}]{Barnes:1986}
Barnes, J., \& {Hut}, P. 1986, \nat, 324, 446

\bibitem[{Baugh {et~al.}(2005)Baugh, Lacey, Frenk, Granato, Silva, Bressan,
  Benson, \& Cole}]{Baugh:2005}
Baugh, C.~M., Lacey, C.~G., Frenk, C.~S., Granato, G.~L., Silva, L., Bressan,
  A., Benson, A.~J., \& Cole, S. 2005, \mnras, 356, 1191

\bibitem[{Biggs \& Ivison(2008)}]{Biggs:2008}
Biggs, A.~D., \& Ivison, R.~J. 2008, \mnras, 385, 893

\bibitem[{Blain {et~al.}(2002)Blain, Smail, Ivison, Kneib, \&
  Frayer}]{Blain:2002}
Blain, A.~W., Smail, I., Ivison, R.~J., Kneib, J.-P., \& Frayer, D.~T. 2002,
  \physrep, 369, 111

\bibitem[{{Bothwell} {et~al.}(2010){Bothwell}, {Chapman}, {Tacconi}, {Smail},
  {Ivison}, {Casey}, {Bertoldi}, {Beswick}, {Biggs}, {Blain}, {Cox}, {Genzel},
  {Greve}, {Kennicutt}, {Muxlow}, {Neri}, \& {Omont}}]{Bothwell:2010}
{Bothwell}, M.~S., {et~al.} 2010, \mnras, 405, 219

\bibitem[{Bouch{\'e} {et~al.}(2007)Bouch{\'e}, Cresci, Davies, Eisenhauer,
  Schreiber, Genzel, Gillessen, Lehnert, Lutz, Nesvadba, Shapiro, Sternberg,
  Tacconi, Verma, Cimatti, Daddi, Renzini, Erb, Shapley, \&
  Steidel}]{Bouche:2007}
Bouch{\'e}, N., {et~al.} 2007, \apj, 671, 303

\bibitem[{{Brodwin} {et~al.}(2008){Brodwin}, {Dey}, {Brown}, {Pope}, {Armus},
  {Bussmann}, {Desai}, {Jannuzi}, \& {Le Floc'h}}]{Brodwin:2008}
{Brodwin}, M., {et~al.} 2008, \apjl, 687, L65

\bibitem[{Bush {et~al.}(2010)Bush, Cox, Hayward, Thilker, Hernquist, \&
  Besla}]{Bush:2010}
Bush, S.~J., Cox, T.~J., Hayward, C.~C., Thilker, D., Hernquist, L., \& Besla,
  G. 2010, \apj, 713, 780

\bibitem[{{Bussmann} {et~al.}(2009{\natexlab{a}}){Bussmann}, {Dey}, {Lotz},
  {Armus}, {Brand}, {Brown}, {Desai}, {Eisenhardt}, {Higdon}, {Higdon},
  {Jannuzi}, {Le Floc'h}, {Melbourne}, {Soifer}, \&
  {Weedman}}]{Bussmann:2009morphology}
{Bussmann}, R.~S., {et~al.} 2009{\natexlab{a}}, \apj, 693, 750

\bibitem[{{Bussmann} {et~al.}(2009{\natexlab{b}}){Bussmann}, {Dey}, {Borys},
  {Desai}, {Jannuzi}, {Le Floc'h}, {Melbourne}, {Sheth}, \&
  {Soifer}}]{Bussmann:2009dog}
---. 2009{\natexlab{b}}, \apj, 705, 184

\bibitem[{Calzetti(1997)}]{Calzetti:1997}
Calzetti, D. 1997, \aj, 113, 162

\bibitem[{Calzetti {et~al.}(2000)Calzetti, Armus, Bohlin, Kinney, Koornneef, \&
  Storchi-Bergmann}]{Calzetti:2000}
Calzetti, D., Armus, L., Bohlin, R.~C., Kinney, A.~L., Koornneef, J., \&
  Storchi-Bergmann, T. 2000, \apj, 533, 682

\bibitem[{Calzetti {et~al.}(1994)Calzetti, Kinney, \&
  Storchi-Bergmann}]{Calzetti:1994}
Calzetti, D., Kinney, A.~L., \& Storchi-Bergmann, T. 1994, \apj, 429, 582

\bibitem[{Capak {et~al.}(2008)Capak, Carilli, Lee, Aldcroft, Aussel,
  Schinnerer, Wilson, Yun, Blain, Giavalisco, Ilbert, Kartaltepe, Lee,
  McCracken, Mobasher, Salvato, Sasaki, Scott, Sheth, Shioya, Thompson, Elvis,
  Sanders, Scoville, \& Tanaguchi}]{Capak:2008}
Capak, P., {et~al.} 2008, \apjl, 681, L53

\bibitem[{{Carilli} {et~al.}(2010){Carilli}, {Daddi}, {Riechers}, {Walter},
  {Weiss}, {Dannerbauer}, {Morrison}, {Wagg}, {Dav{\'e}}, {Elbaz}, {Stern},
  {Dickinson}, {Krips}, \& {Aravena}}]{Carilli:2010}
{Carilli}, C.~L., {et~al.} 2010, \apj, 714, 1407

\bibitem[{{Casey} {et~al.}(2010){Casey}, {Chapman}, {Smail}, {Alaghband-Zadeh},
  {Bothwell}, \& {Swinbank}}]{Casey:2010}
{Casey}, C.~M., {Chapman}, S.~C., {Smail}, I., {Alaghband-Zadeh}, S.,
  {Bothwell}, M.~S., \& {Swinbank}, A.~M. 2010, arXiv:1009.5709

\bibitem[{Casey {et~al.}(2009)Casey, Chapman, Beswick, Biggs, Blain, Hainline,
  Ivison, Muxlow, \& Smail}]{Casey:2009}
Casey, C.~M., {et~al.} 2009, \mnras, 399, 121

\bibitem[{Chakrabarti {et~al.}(2007)Chakrabarti, Cox, Hernquist, Hopkins,
  Robertson, \& Matteo}]{Chakrabarti:2007}
Chakrabarti, S., Cox, T.~J., Hernquist, L., Hopkins, P.~F., Robertson, B., \&
  Matteo, T.~D. 2007, \apj, 658, 840

\bibitem[{Chakrabarti {et~al.}(2008)Chakrabarti, Fenner, Cox, Hernquist, \&
  Whitney}]{Chakrabarti:2008SMG}
Chakrabarti, S., Fenner, Y., Cox, T.~J., Hernquist, L., \& Whitney, B.~A. 2008,
  \apj, 688, 972

\bibitem[{{Chakrabarti} \& {McKee}(2005)}]{Chakrabarti:2005RT}
{Chakrabarti}, S., \& {McKee}, C.~F. 2005, \apj, 631, 792

\bibitem[{{Chakrabarti} \& {McKee}(2008)}]{Chakrabarti:2008RT}
---. 2008, \apj, 683, 693

\bibitem[{Chakrabarti \& Whitney(2009)}]{Chakrabarti:2009}
Chakrabarti, S., \& Whitney, B.~A. 2009, \apj, 690, 1432

\bibitem[{Chapman {et~al.}(2005)Chapman, Blain, Smail, \&
  Ivison}]{Chapman:2005}
Chapman, S.~C., Blain, A.~W., Smail, I., \& Ivison, R.~J. 2005, \apj, 622, 772

\bibitem[{Chapman {et~al.}(2004)Chapman, Smail, Blain, \&
  Ivison}]{Chapman:2004}
Chapman, S.~C., Smail, I., Blain, A.~W., \& Ivison, R.~J. 2004, \apj, 614, 671

\bibitem[{{Chapman} {et~al.}(2003){Chapman}, {Windhorst}, {Odewahn}, {Yan}, \&
  {Conselice}}]{Chapman:2003}
{Chapman}, S.~C., {Windhorst}, R., {Odewahn}, S., {Yan}, H., \& {Conselice}, C.
  2003, \apj, 599, 92

\bibitem[{Chapman {et~al.}(2000)Chapman, Scott, Steidel, Borys, Halpern,
  Morris, Adelberger, Dickinson, Giavalisco, \& Pettini}]{Chapman:2000}
Chapman, S.~C., {et~al.} 2000, \mnras, 319, 318

\bibitem[{{Chapman} {et~al.}(2010){Chapman}, {Ivison}, {Roseboom}, {Auld},
  {Bock}, {Brisbin}, {Burgarella}, {Chanial}, {Clements}, {Cooray}, {Eales},
  {Franceschini}, {Giovannoli}, {Glenn}, {Griffin}, {Mortier}, {Oliver},
  {Omont}, {Page}, {Papageorgiou}, {Pearson}, {P{\'e}rez-Fournon}, {Pohlen},
  {Rawlings}, {Raymond}, {Rodighiero}, {Rowan-Robinson}, {Scott}, {Seymour},
  {Smith}, {Symeonidis}, {Tugwell}, {Vaccari}, {Vieira}, {Vigroux}, {Wang}, \&
  {Wright}}]{Chapman:2010}
{Chapman}, S.~C., {et~al.} 2010, \mnras, 409, L13

\bibitem[{Chary \& Elbaz(2001)}]{Chary:2001}
Chary, R., \& Elbaz, D. 2001, \apj, 556, 562

\bibitem[{Clements {et~al.}(2010)Clements, Dunne, \& Eales}]{Clements:2010}
Clements, D.~L., Dunne, L., \& Eales, S. 2010, \mnras, 403, 274

\bibitem[{{Conley} {et~al.}(2011){Conley}, {Cooray}, {Vieira}, {Gonz{\'a}lez
  Solares}, {Kim}, {Aguirre}, {Amblard}, {Auld}, {Baker}, {Beelen}, {Blain},
  {Blundell}, {Bock}, {Bradford}, {Bridge}, {Brisbin}, {Burgarella},
  {Carpenter}, {Chanial}, {Chapin}, {Christopher}, {Clements}, {Cox},
  {Djorgovski}, {Dowell}, {Eales}, {Earle}, {Ellsworth-Bowers}, {Farrah},
  {Franceschini}, {Frayer}, {Fu}, {Gavazzi}, {Glenn}, {Griffin}, {Gurwell},
  {Halpern}, {Ibar}, {Ivison}, {Jarvis}, {Kamenetzky}, {Krips}, {Levenson},
  {Lupu}, {Mahabal}, {Maloney}, {Maraston}, {Marchetti}, {Marsden},
  {Matsuhara}, {Mortier}, {Murphy}, {Naylor}, {Neri}, {Nguyen}, {Oliver},
  {Omont}, {Page}, {Papageorgiou}, {Pearson}, {P{\'e}rez-Fournon}, {Pohlen},
  {Rangwala}, {Rawlings}, {Raymond}, {Riechers}, {Rodighiero}, {Roseboom},
  {Rowan-Robinson}, {Schulz}, {Scott}, {Scott}, {Serra}, {Seymour}, {Shupe},
  {Smith}, {Symeonidis}, {Tugwell}, {Vaccari}, {Valiante}, {Valtchanov},
  {Verma}, {Viero}, {Vigroux}, {Wang}, {Wiebe}, {Wright}, {Xu}, {Zeimann},
  {Zemcov}, \& {Zmuidzinas}}]{Conley:2011}
{Conley}, A., {et~al.} 2011, \apjl, 732, L35

\bibitem[{Coppin {et~al.}(2008)Coppin, Halpern, Scott, Borys, Dunlop, Dunne,
  Ivison, Wagg, Aretxaga, Battistelli, Benson, Blain, Chapman, Clements, Dye,
  Farrah, Hughes, Jenness, van Kampen, Lacey, Mortier, Pope, Priddey, Serjeant,
  Smail, Stevens, \& Vaccari}]{Coppin:2008}
Coppin, K., {et~al.} 2008, \mnras, 384, 1597

\bibitem[{{Cox} {et~al.}(2006{\natexlab{a}}){Cox}, Dutta, Matteo, Hernquist,
  Hopkins, Robertson, \& Springel}]{Cox:2006}
{Cox}, T.~J., Dutta, S.~N., Matteo, T.~D., Hernquist, L., Hopkins, P.~F.,
  Robertson, B., \& Springel, V. 2006{\natexlab{a}}, \apj, 650, 791

\bibitem[{{Cox} {et~al.}(2006{\natexlab{b}}){Cox}, {Jonsson}, {Primack}, \&
  {Somerville}}]{Cox:2006feedback}
{Cox}, T.~J., {Jonsson}, P., {Primack}, J.~R., \& {Somerville}, R.~S.
  2006{\natexlab{b}}, \mnras, 373, 1013

\bibitem[{{Cox} {et~al.}(2008){Cox}, {Jonsson}, {Somerville}, {Primack}, \&
  {Dekel}}]{Cox:2008}
{Cox}, T.~J., {Jonsson}, P., {Somerville}, R.~S., {Primack}, J.~R., \& {Dekel},
  A. 2008, \mnras, 384, 386

\bibitem[{Daddi {et~al.}(2009{\natexlab{a}})Daddi, Dannerbauer, Krips, Walter,
  Dickinson, Elbaz, \& Morrison}]{Daddi:2009smga}
Daddi, E., Dannerbauer, H., Krips, M., Walter, F., Dickinson, M., Elbaz, D., \&
  Morrison, G.~E. 2009{\natexlab{a}}, \apjl, 695, L176

\bibitem[{Daddi {et~al.}(2005)Daddi, {Dickinson}, {Chary}, {Pope}, {Morrison},
  {Alexander}, {Bauer}, {Brandt}, {Giavalisco}, {Ferguson}, {Lee}, {Lehmer},
  {Papovich}, \& {Renzini}}]{Daddi:2005}
Daddi, E., {et~al.} 2005, \apjl, 631, L13

\bibitem[{Daddi {et~al.}(2007)Daddi, Dickinson, Morrison, Chary, Cimatti,
  Elbaz, Frayer, Renzini, Pope, Alexander, Bauer, Giavalisco, Huynh, Kurk, \&
  Mignoli}]{Daddi:2007}
---. 2007, \apj, 670, 156

\bibitem[{Daddi {et~al.}(2009{\natexlab{b}})Daddi, Dannerbauer, Stern,
  Dickinson, Morrison, Elbaz, Giavalisco, Mancini, Pope, \&
  Spinrad}]{Daddi:2009smgb}
---. 2009{\natexlab{b}}, \apj, 694, 1517

\bibitem[{Daddi {et~al.}(2010)Daddi, {Bournaud}, {Walter}, {Dannerbauer},
  {Carilli}, {Dickinson}, {Elbaz}, {Morrison}, {Riechers}, {Onodera}, {Salmi},
  {Krips}, \& {Stern}}]{Daddi:2010}
---. 2010, \apj, 713, 686

\bibitem[{Dale \& Helou(2002)}]{Dale:2002}
Dale, D.~A., \& Helou, G. 2002, \apj, 576, 159

\bibitem[{Dale {et~al.}(2001)Dale, Helou, Contursi, Silbermann, \&
  Kolhatkar}]{Dale:2001}
Dale, D.~A., Helou, G., Contursi, A., Silbermann, N.~A., \& Kolhatkar, S. 2001,
  \apj, 549, 215

\bibitem[{{Dale} {et~al.}(2007){Dale}, {Gil de Paz}, {Gordon}, {Hanson},
  {Armus}, {Bendo}, {Bianchi}, {Block}, {Boissier}, {Boselli}, {Buckalew},
  {Buat}, {Burgarella}, {Calzetti}, {Cannon}, {Engelbracht}, {Helou},
  {Hollenbach}, {Jarrett}, {Kennicutt}, {Leitherer}, {Li}, {Madore}, {Martin},
  {Meyer}, {Murphy}, {Regan}, {Roussel}, {Smith}, {Sosey}, {Thilker}, \&
  {Walter}}]{Dale:2007}
{Dale}, D.~A., {et~al.} 2007, \apj, 655, 863

\bibitem[{{Dannerbauer} {et~al.}(2009){Dannerbauer}, {Daddi}, {Riechers},
  {Walter}, {Carilli}, {Dickinson}, {Elbaz}, \& {Morrison}}]{Dannerbauer:2009}
{Dannerbauer}, H., {Daddi}, E., {Riechers}, D.~A., {Walter}, F., {Carilli},
  C.~L., {Dickinson}, M., {Elbaz}, D., \& {Morrison}, G.~E. 2009, \apjl, 698,
  L178

\bibitem[{{Dannerbauer} {et~al.}(2002){Dannerbauer}, {Lehnert}, {Lutz},
  {Tacconi}, {Bertoldi}, {Carilli}, {Genzel}, \& {Menten}}]{Dannerbauer:2002}
{Dannerbauer}, H., {Lehnert}, M.~D., {Lutz}, D., {Tacconi}, L., {Bertoldi}, F.,
  {Carilli}, C., {Genzel}, R., \& {Menten}, K. 2002, \apj, 573, 473

\bibitem[{{Dannerbauer} {et~al.}(2010){Dannerbauer}, {Daddi}, {Morrison},
  {Altieri}, {Andreani}, {Aussel}, {Berta}, {Bongiovanni}, {Cava}, {Cepa},
  {Cimatti}, {Dominguez}, {Elbaz}, {F{\"o}rster Schreiber}, {Genzel},
  {Gruppioni}, {Horeau}, {Hwang}, {Le Floc'h}, {Le Pennec}, {Lutz}, {Magdis},
  {Magnelli}, {Maiolino}, {Nordon}, {P{\'e}rez Garc{\'{\i}}a}, {Poglitsch},
  {Popesso}, {Pozzi}, {Riguccini}, {Rodighiero}, {Saintonge}, {Santini},
  {Sanchez-Portal}, {Shao}, {Sturm}, {Tacconi}, \&
  {Valtchanov}}]{Dannerbauer:2010}
{Dannerbauer}, H., {et~al.} 2010, \apjl, 720, L144

\bibitem[{Dav{\'e} {et~al.}(2010)Dav{\'e}, Finlator, Oppenheimer, Fardal, Katz,
  Kere{\v s}, \& Weinberg}]{Dave:2010}
Dav{\'e}, R., Finlator, K., Oppenheimer, B.~D., Fardal, M., Katz, N., Kere{\v
  s}, D., \& Weinberg, D.~H. 2010, \mnras, 404, 1355

\bibitem[{{Dekel} {et~al.}(2009){Dekel}, {Birnboim}, {Engel}, {Freundlich},
  {Goerdt}, {Mumcuoglu}, {Neistein}, {Pichon}, {Teyssier}, \&
  {Zinger}}]{Dekel:2009nature}
{Dekel}, A., {et~al.} 2009, \nat, 457, 451

\bibitem[{Downes \& Solomon(1998)}]{Downes:1998}
Downes, D., \& Solomon, P.~M. 1998, \apj, 507, 615

\bibitem[{Draine \& Li(2007)}]{Draine:2007}
Draine, B.~T., \& Li, A. 2007, \apj, 657, 810

\bibitem[{{Dunne} \& {Eales}(2001)}]{Dunne:2001}
{Dunne}, L., \& {Eales}, S.~A. 2001, \mnras, 327, 697

\bibitem[{{Dunne} {et~al.}(2003){Dunne}, {Eales}, \& {Edmunds}}]{Dunne:2003}
{Dunne}, L., {Eales}, S.~A., \& {Edmunds}, M.~G. 2003, \mnras, 341, 589

\bibitem[{Dwek(1998)}]{Dwek:1998}
Dwek, E. 1998, \apj, 501, 643

\bibitem[{Eales {et~al.}(1999)Eales, Lilly, Gear, Dunne, Bond, Hammer,
  F{\`e}vre, \& Crampton}]{Eales:1999}
Eales, S., Lilly, S., Gear, W., Dunne, L., Bond, J.~R., Hammer, F., F{\`e}vre,
  O.~L., \& Crampton, D. 1999, \apj, 515, 518

\bibitem[{Edmunds \& Eales(1998)}]{Edmunds:1998}
Edmunds, M.~G., \& Eales, S.~A. 1998, \mnras, 299, L29

\bibitem[{{Engel} {et~al.}(2010){Engel}, {Tacconi}, {Davies}, {Neri}, {Smail},
  {Chapman}, {Genzel}, {Cox}, {Greve}, {Ivison}, {Blain}, {Bertoldi}, \&
  {Omont}}]{Engel:2010}
{Engel}, H., {et~al.} 2010, arXiv:1009.2495

\bibitem[{{Erb} {et~al.}(2006){Erb}, {Steidel}, {Shapley}, {Pettini}, {Reddy},
  \& {Adelberger}}]{Erb:2006}
{Erb}, D.~K., {Steidel}, C.~C., {Shapley}, A.~E., {Pettini}, M., {Reddy},
  N.~A., \& {Adelberger}, K.~L. 2006, \apj, 646, 107

\bibitem[{{Fontanot} \& {Monaco}(2010)}]{Fontanot:2010}
{Fontanot}, F., \& {Monaco}, P. 2010, \mnras, 405, 705

\bibitem[{{Fontanot} {et~al.}(2007){Fontanot}, {Monaco}, {Silva}, \&
  {Grazian}}]{Fontanot:2007}
{Fontanot}, F., {Monaco}, P., {Silva}, L., \& {Grazian}, A. 2007, \mnras, 382,
  903

\bibitem[{{Gingold} \& {Monaghan}(1977)}]{Gingold:1977}
{Gingold}, R.~A., \& {Monaghan}, J.~J. 1977, \mnras, 181, 375

\bibitem[{{Gonz{\'a}lez} {et~al.}(2011){Gonz{\'a}lez}, {Lacey}, {Baugh}, \&
  {Frenk}}]{Gonzalez:2011}
{Gonz{\'a}lez}, J.~E., {Lacey}, C.~G., {Baugh}, C.~M., \& {Frenk}, C.~S. 2011,
  \mnras, 413, 749

\bibitem[{{Granato} {et~al.}(2004){Granato}, {De Zotti}, {Silva}, {Bressan}, \&
  {Danese}}]{Granato:2004}
{Granato}, G.~L., {De Zotti}, G., {Silva}, L., {Bressan}, A., \& {Danese}, L.
  2004, \apj, 600, 580

\bibitem[{{Greve} {et~al.}(2008){Greve}, {Pope}, {Scott}, {Ivison}, {Borys},
  {Conselice}, \& {Bertoldi}}]{Greve:2008}
{Greve}, T.~R., {Pope}, A., {Scott}, D., {Ivison}, R.~J., {Borys}, C.,
  {Conselice}, C.~J., \& {Bertoldi}, F. 2008, \mnras, 389, 1489

\bibitem[{Greve {et~al.}(2005)Greve, Bertoldi, Smail, Neri, Chapman, Blain,
  Ivison, Genzel, Omont, Cox, Tacconi, \& Kneib}]{Greve:2005}
Greve, T.~R., {et~al.} 2005, \mnras, 359, 1165

\bibitem[{Groves {et~al.}(2008)Groves, Dopita, Sutherland, Kewley, Fischera,
  Leitherer, Brandl, \& van Breugel}]{Groves:2008}
Groves, B., Dopita, M.~A., Sutherland, R.~S., Kewley, L.~J., Fischera, J.,
  Leitherer, C., Brandl, B., \& van Breugel, W. 2008, \apjs, 176, 438

\bibitem[{{Hainline} {et~al.}(2010){Hainline}, {Blain}, {Smail}, {Alexander},
  {Armus}, {Chapman}, \& {Ivison}}]{Hainline:2010}
{Hainline}, L.~J., {Blain}, A.~W., {Smail}, I., {Alexander}, D.~M., {Armus},
  L., {Chapman}, S.~C., \& {Ivison}, R.~J. 2010, arXiv:1006.0238

\bibitem[{{Hayward} {et~al.}(2011){Hayward}, {Narayanan}, {Jonsson}, {Cox},
  {Kere{\v s}}, {Hopkins}, \& {Hernquist}}]{Hayward:2011num_cts_proc}
{Hayward}, C.~C., {Narayanan}, D., {Jonsson}, P., {Cox}, T.~J., {Kere{\v s}},
  D., {Hopkins}, P.~F., \& {Hernquist}, L. 2011, in Astronomical Society of the
  Pacific Conference Series, Vol. 440, Astronomical Society of the Pacific
  Conference Series, ed. {M.~Treyer, T.~Wyder, J.~Neill, M.~Seibert, \&
  J.~Lee}, 369--+

\bibitem[{Hernquist(1990)}]{Hernquist:1990}
Hernquist, L. 1990, \apj, 356, 359

\bibitem[{{Hernquist} \& {Katz}(1989)}]{Hernquist:1989treesph}
{Hernquist}, L., \& {Katz}, N. 1989, \apjs, 70, 419

\bibitem[{Hobson \& Padman(1993)}]{Hobson:1993}
Hobson, M.~P., \& Padman, R. 1993, \mnras, 264, L161

\bibitem[{Holland {et~al.}(1999)Holland, Robson, Gear, Cunningham, Lightfoot,
  Jenness, Ivison, Stevens, Ade, Griffin, Duncan, Murphy, \&
  Naylor}]{Holland:1999}
Holland, W.~S., {et~al.} 1999, \mnras, 303, 659

\bibitem[{{Hopkins} {et~al.}(2009{\natexlab{a}}){Hopkins}, {Cox}, {Dutta},
  {Hernquist}, {Kormendy}, \& {Lauer}}]{Hopkins:2009cusps}
{Hopkins}, P.~F., {Cox}, T.~J., {Dutta}, S.~N., {Hernquist}, L., {Kormendy},
  J., \& {Lauer}, T.~R. 2009{\natexlab{a}}, \apjs, 181, 135

\bibitem[{{Hopkins} {et~al.}(2008{\natexlab{a}}){Hopkins}, {Cox}, {Kere{\v s}},
  \& {Hernquist}}]{Hopkins:2008red_Es}
{Hopkins}, P.~F., {Cox}, T.~J., {Kere{\v s}}, D., \& {Hernquist}, L.
  2008{\natexlab{a}}, \apjs, 175, 390

\bibitem[{{Hopkins} {et~al.}(2009{\natexlab{b}}){Hopkins}, Cox, Younger, \&
  Hernquist}]{Hopkins:2009disk_survival}
{Hopkins}, P.~F., Cox, T.~J., Younger, J.~D., \& Hernquist, L.
  2009{\natexlab{b}}, \apj, 691, 1168

\bibitem[{{Hopkins} \& {Hernquist}(2010)}]{HH:2010}
{Hopkins}, P.~F., \& {Hernquist}, L. 2010, \mnras, 402, 985

\bibitem[{{Hopkins} {et~al.}(2006{\natexlab{a}}){Hopkins}, {Hernquist}, {Cox},
  {Di Matteo}, {Robertson}, \& {Springel}}]{Hopkins:2006unified_model}
{Hopkins}, P.~F., {Hernquist}, L., {Cox}, T.~J., {Di Matteo}, T., {Robertson},
  B., \& {Springel}, V. 2006{\natexlab{a}}, \apjs, 163, 1

\bibitem[{{Hopkins} {et~al.}(2008{\natexlab{b}}){Hopkins}, {Hernquist}, {Cox},
  {Dutta}, \& {Rothberg}}]{Hopkins:2008extra_light}
{Hopkins}, P.~F., {Hernquist}, L., {Cox}, T.~J., {Dutta}, S.~N., \& {Rothberg},
  B. 2008{\natexlab{b}}, \apj, 679, 156

\bibitem[{{Hopkins} {et~al.}(2008{\natexlab{c}}){Hopkins}, {Hernquist}, {Cox},
  \& {Kere{\v s}}}]{Hopkins:2008cosm_frame1}
{Hopkins}, P.~F., {Hernquist}, L., {Cox}, T.~J., \& {Kere{\v s}}, D.
  2008{\natexlab{c}}, \apjs, 175, 356

\bibitem[{{Hopkins} {et~al.}(2006{\natexlab{b}}){Hopkins}, {Hernquist}, {Cox},
  {Robertson}, \& {Springel}}]{Hopkins:2006quasar_LF}
{Hopkins}, P.~F., {Hernquist}, L., {Cox}, T.~J., {Robertson}, B., \&
  {Springel}, V. 2006{\natexlab{b}}, \apjs, 163, 50

\bibitem[{{Hopkins} {et~al.}(2009{\natexlab{c}}){Hopkins}, {Lauer}, {Cox},
  {Hernquist}, \& {Kormendy}}]{Hopkins:2009cores}
{Hopkins}, P.~F., {Lauer}, T.~R., {Cox}, T.~J., {Hernquist}, L., \& {Kormendy},
  J. 2009{\natexlab{c}}, \apjs, 181, 486

\bibitem[{{Hopkins} {et~al.}(2011){Hopkins}, {Quataert}, \&
  {Murray}}]{Hopkins:2011self-regulated_SF}
{Hopkins}, P.~F., {Quataert}, E., \& {Murray}, N. 2011, arXiv:1101.4940

\bibitem[{{Hopkins} {et~al.}(2007){Hopkins}, Richards, \&
  Hernquist}]{Hopkins:2007}
{Hopkins}, P.~F., Richards, G.~T., \& Hernquist, L. 2007, \apj, 654, 731

\bibitem[{{Hopkins} {et~al.}(2006{\natexlab{c}}){Hopkins}, {Somerville},
  {Hernquist}, {Cox}, {Robertson}, \& {Li}}]{Hopkins:2006LFs_SFH}
{Hopkins}, P.~F., {Somerville}, R.~S., {Hernquist}, L., {Cox}, T.~J.,
  {Robertson}, B., \& {Li}, Y. 2006{\natexlab{c}}, \apj, 652, 864

\bibitem[{{Hopkins} {et~al.}(2010){Hopkins}, Younger, Hayward, Narayanan, \&
  Hernquist}]{Hopkins:2010IR_LF}
{Hopkins}, P.~F., Younger, J.~D., Hayward, C.~C., Narayanan, D., \& Hernquist,
  L. 2010, \mnras, 402, 1693

\bibitem[{{Hopkins} {et~al.}(2009{\natexlab{d}}){Hopkins}, {Somerville}, {Cox},
  {Hernquist}, {Jogee}, {Kere{\v s}}, {Ma}, {Robertson}, \&
  {Stewart}}]{Hopkins:2009bulge_demo}
{Hopkins}, P.~F., {et~al.} 2009{\natexlab{d}}, \mnras, 397, 802

\bibitem[{Hughes {et~al.}(1998)Hughes, Serjeant, Dunlop, Rowan-Robinson, Blain,
  Mann, Ivison, Peacock, Efstathiou, Gear, Oliver, Lawrence, Longair,
  Goldschmidt, \& Jenness}]{Hughes:1998}
Hughes, D.~H., {et~al.} 1998, \nat, 394, 241

\bibitem[{{Hwang} {et~al.}(2010){Hwang}, {Elbaz}, {Magdis}, {Daddi},
  {Symeonidis}, {Altieri}, {Amblard}, {Andreani}, {Arumugam}, {Auld}, {Aussel},
  {Babbedge}, {Berta}, {Blain}, {Bock}, {Bongiovanni}, {Boselli}, {Buat},
  {Burgarella}, {Castro-Rodr{\'{\i}}guez}, {Cava}, {Cepa}, {Chanial}, {Chapin},
  {Chary}, {Cimatti}, {Clements}, {Conley}, {Conversi}, {Cooray},
  {Dannerbauer}, {Dickinson}, {Dominguez}, {Dowell}, {Dunlop}, {Dwek}, {Eales},
  {Farrah}, {Schreiber}, {Fox}, {Franceschini}, {Gear}, {Genzel}, {Glenn},
  {Griffin}, {Gruppioni}, {Halpern}, {Hatziminaoglou}, {Ibar}, {Isaak},
  {Ivison}, {Jeong}, {Lagache}, {Le Borgne}, {Le Floc'h}, {Lee}, {Lee}, {Lee},
  {Levenson}, {Lu}, {Lutz}, {Madden}, {Maffei}, {Magnelli}, {Mainetti},
  {Maiolino}, {Marchetti}, {Mortier}, {Nguyen}, {Nordon}, {O'Halloran},
  {Okumura}, {Oliver}, {Omont}, {Page}, {Panuzzo}, {Papageorgiou}, {Pearson},
  {P{\'e}rez-Fournon}, {Garc{\'{\i}}a}, {Poglitsch}, {Pohlen}, {Popesso},
  {Pozzi}, {Rawlings}, {Rigopoulou}, {Riguccini}, {Rizzo}, {Rodighiero},
  {Roseboom}, {Rowan-Robinson}, {Saintonge}, {Portal}, {Santini}, {Sauvage},
  {Schulz}, {Scott}, {Seymour}, {Shao}, {Shupe}, {Smith}, {Stevens}, {Sturm},
  {Tacconi}, {Trichas}, {Tugwell}, {Vaccari}, {Valtchanov}, {Vieira},
  {Vigroux}, {Wang}, {Ward}, {Wright}, {Xu}, \&
  {Zemcov}}]{Hwang:2010dust_T_evolution}
{Hwang}, H.~S., {et~al.} 2010, \mnras, 409, 75

\bibitem[{Iono {et~al.}(2009)Iono, Wilson, Yun, Baker, Petitpas, Peck, Krips,
  Cox, Matsushita, Mihos, \& Pihlstrom}]{Iono:2009}
Iono, D., {et~al.} 2009, \apj, 695, 1537

\bibitem[{{Ivison} {et~al.}(2010){Ivison}, {Smail}, {Papadopoulos}, {Wold},
  {Richard}, {Swinbank}, {Kneib}, \& {Owen}}]{Ivison:2010}
{Ivison}, R.~J., {Smail}, I., {Papadopoulos}, P.~P., {Wold}, I., {Richard}, J.,
  {Swinbank}, A.~M., {Kneib}, J., \& {Owen}, F.~N. 2010, \mnras, 404, 198

\bibitem[{{Ivison} {et~al.}(2002){Ivison}, {Greve}, {Smail}, {Dunlop}, {Roche},
  {Scott}, {Page}, {Stevens}, {Almaini}, {Blain}, {Willott}, {Fox}, {Gilbank},
  {Serjeant}, \& {Hughes}}]{Ivison:2002}
{Ivison}, R.~J., {et~al.} 2002, \mnras, 337, 1

\bibitem[{{Ivison} {et~al.}(2007){Ivison}, Greve, Dunlop, Peacock, Egami,
  Smail, Ibar, van Kampen, Aretxaga, Babbedge, Biggs, Blain, Chapman, Clements,
  Coppin, Farrah, Halpern, Hughes, Jarvis, Jenness, Jones, Mortier, Oliver,
  Papovich, P{\'e}rez-Gonz{\'a}lez, Pope, Rawlings, Rieke, Rowan-Robinson,
  Savage, Scott, Seigar, Serjeant, Simpson, Stevens, Vaccari, Wagg, \&
  Willott}]{Ivison:2007}
---. 2007, \mnras, 380, 199

\bibitem[{James {et~al.}(2002)James, Dunne, Eales, \& Edmunds}]{James:2002}
James, A., Dunne, L., Eales, S., \& Edmunds, M.~G. 2002, \mnras, 335, 753

\bibitem[{Jonsson(2006)}]{Jonsson:2006sunrise}
Jonsson, P. 2006, \mnras, 372, 2

\bibitem[{Jonsson {et~al.}(2006)Jonsson, Cox, Primack, \&
  Somerville}]{Jonsson:2006}
Jonsson, P., Cox, T.~J., Primack, J.~R., \& Somerville, R.~S. 2006, \apj, 637,
  255

\bibitem[{Jonsson {et~al.}(2010)Jonsson, Groves, \& Cox}]{Jonsson:2010sunrise}
Jonsson, P., Groves, B.~A., \& Cox, T.~J. 2010, \mnras, 403, 17

\bibitem[{Jonsson \& Primack(2010)}]{Jonsson:2010gpu}
Jonsson, P., \& Primack, J.~R. 2010, NewA, 15, 509

\bibitem[{Juvela(2005)}]{Juvela:2005}
Juvela, M. 2005, \aap, 440, 531

\bibitem[{Katz {et~al.}(1996)Katz, Weinberg, \& Hernquist}]{Katz:1996}
Katz, N., Weinberg, D.~H., \& Hernquist, L. 1996, \apjs, 105, 19

\bibitem[{Kennicutt(1998{\natexlab{a}})}]{Kennicutt:1998}
Kennicutt, R.~C. 1998{\natexlab{a}}, \apj, 498, 541

\bibitem[{Kennicutt(1998{\natexlab{b}})}]{Kennicutt:1998review}
---. 1998{\natexlab{b}}, \araa, 36, 189

\bibitem[{{Kennicutt} {et~al.}(2003){Kennicutt}, {Armus}, {Bendo}, {Calzetti},
  {Dale}, {Draine}, {Engelbracht}, {Gordon}, {Grauer}, {Helou}, {Hollenbach},
  {Jarrett}, {Kewley}, {Leitherer}, {Li}, {Malhotra}, {Regan}, {Rieke},
  {Rieke}, {Roussel}, {Smith}, {Thornley}, \& {Walter}}]{Kennicutt:2003}
{Kennicutt}, Jr., R.~C., {et~al.} 2003, \pasp, 115, 928

\bibitem[{Kere{\v s} {et~al.}(2009)Kere{\v s}, Katz, Fardal, Dav{\'e}, \&
  Weinberg}]{Keres:2009a}
Kere{\v s}, D., Katz, N., Fardal, M., Dav{\'e}, R., \& Weinberg, D.~H. 2009,
  \mnras, 395, 160

\bibitem[{Kere{\v s} {et~al.}(2005)Kere{\v s}, Katz, Weinberg, \&
  Dav{\'e}}]{Keres:2005}
Kere{\v s}, D., Katz, N., Weinberg, D.~H., \& Dav{\'e}, R. 2005, \mnras, 363, 2

\bibitem[{Knudsen {et~al.}(2010)Knudsen, Kneib, Richard, Petitpas, \&
  Egami}]{Knudsen:2010}
Knudsen, K.~K., Kneib, J.-P., Richard, J., Petitpas, G., \& Egami, E. 2010,
  \apj, 709, 210

\bibitem[{{Kormendy} {et~al.}(2009){Kormendy}, {Fisher}, {Cornell}, \&
  {Bender}}]{Kormendy:2009}
{Kormendy}, J., {Fisher}, D.~B., {Cornell}, M.~E., \& {Bender}, R. 2009, \apjs,
  182, 216

\bibitem[{{Kov{\'a}cs} {et~al.}(2006){Kov{\'a}cs}, Chapman, Dowell, Blain,
  Ivison, Smail, \& Phillips}]{Kovacs:2006}
{Kov{\'a}cs}, A., Chapman, S.~C., Dowell, C.~D., Blain, A.~W., Ivison, R.~J.,
  Smail, I., \& Phillips, T.~G. 2006, \apj, 650, 592

\bibitem[{{Kov{\'a}cs} {et~al.}(2010){Kov{\'a}cs}, {Omont}, {Beelen},
  {Lonsdale}, {Polletta}, {Fiolet}, {Greve}, {Borys}, {Cox}, {De Breuck},
  {Dole}, {Dowell}, {Farrah}, {Lagache}, {Menten}, {Bell}, \&
  {Owen}}]{Kovacs:2010}
{Kov{\'a}cs}, A., {et~al.} 2010, \apj, 717, 29

\bibitem[{Kroupa(2001)}]{Kroupa:2001}
Kroupa, P. 2001, \mnras, 322, 231

\bibitem[{{Krumholz} \& {Thompson}(2007)}]{Krumholz:2007KS}
{Krumholz}, M.~R., \& {Thompson}, T.~A. 2007, \apj, 669, 289

\bibitem[{Leitherer {et~al.}(1999)Leitherer, Schaerer, Goldader, Delgado,
  Robert, Kune, de~Mello, Devost, \& Heckman}]{Leitherer:1999}
Leitherer, C., {et~al.} 1999, \apjs, 123, 3

\bibitem[{Lisenfeld {et~al.}(2000)Lisenfeld, Isaak, \& Hills}]{Lisenfeld:2000}
Lisenfeld, U., Isaak, K.~G., \& Hills, R. 2000, \mnras, 312, 433

\bibitem[{{Lo Faro} {et~al.}(2009){Lo Faro}, {Monaco}, {Vanzella}, {Fontanot},
  {Silva}, \& {Cristiani}}]{LoFaro:2009}
{Lo Faro}, B., {Monaco}, P., {Vanzella}, E., {Fontanot}, F., {Silva}, L., \&
  {Cristiani}, S. 2009, \mnras, 399, 827

\bibitem[{{Lonsdale} {et~al.}(2006){Lonsdale}, {Farrah}, \&
  {Smith}}]{Lonsdale:2006}
{Lonsdale}, C.~J., {Farrah}, D., \& {Smith}, H.~E. 2006, {Ultraluminous
  Infrared Galaxies}, ed. {Mason, J.~W.} (Springer Verlag), 285

\bibitem[{{Lucy}(1977)}]{Lucy:1977}
{Lucy}, L.~B. 1977, \aj, 82, 1013

\bibitem[{Lupu {et~al.}(2010)Lupu, Scott, Aguirre, Aretxaga, Auld, Barton,
  Beelen, Bertoldi, Bock, Bonfield, Bradford, Buttiglione, Cava, Clements,
  Cooke, Cooray, Dannerbauer, Dariush, Zotti, Dunne, Dye, Eales, Frayer, Fritz,
  Glenn, Hughes, Ibar, Ivison, Jarvis, Kamenetzky, Kim, Lagache, Leeuw, Maddox,
  Maloney, Matsuhara, Murphy, Naylor, Negrello, Nguien, Omont, Pascale, Pohlen,
  Rigby, Rodighiero, Serjeant, Smith, Temi, Thompson, Valtchanov, Verma,
  Vieira, \& Zmuidzinas}]{Lupu:2010}
Lupu, R.~E., {et~al.} 2010, arXiv:1009.5983

\bibitem[{{Magdis} {et~al.}(2010){Magdis}, {Elbaz}, {Hwang}, {Amblard},
  {Arumugam}, {Aussel}, {Blain}, {Bock}, {Boselli}, {Buat},
  {Castro-Rodr{\'{\i}}guez}, {Cava}, {Chanial}, {Clements}, {Conley},
  {Conversi}, {Cooray}, {Dowell}, {Dwek}, {Eales}, {Farrah}, {Franceschini},
  {Glenn}, {Griffin}, {Halpern}, {Hatziminaoglou}, {Huang}, {Ibar}, {Isaak},
  {Le Floc'h}, {Lagache}, {Levenson}, {Lonsdale}, {Lu}, {Madden}, {Maffei},
  {Mainetti}, {Marchetti}, {Morrison}, {Nguyen}, {O'Halloran}, {Oliver},
  {Omont}, {Owen}, {Page}, {Pannella}, {Panuzzo}, {Papageorgiou}, {Pearson},
  {P{\'e}rez-Fournon}, {Pohlen}, {Rigopoulou}, {Rizzo}, {Roseboom},
  {Rowan-Robinson}, {Schulz}, {Scott}, {Seymour}, {Shupe}, {Smith}, {Stevens},
  {Strazzullo}, {Symeonidis}, {Trichas}, {Tugwell}, {Vaccari}, {Valtchanov},
  {Vigroux}, {Wang}, {Wright}, {Xu}, \& {Zemcov}}]{Magdis:2010dust_T}
{Magdis}, G.~E., {et~al.} 2010, \mnras, 409, 22

\bibitem[{{Magnelli} {et~al.}(2010){Magnelli}, {Lutz}, {Berta}, {Altieri},
  {Andreani}, {Aussel}, {Casta{\~n}eda}, {Cava}, {Cepa}, {Cimatti}, {Daddi},
  {Dannerbauer}, {Dominguez}, {Elbaz}, {F{\"o}rster Schreiber}, {Genzel},
  {Grazian}, {Gruppioni}, {Magdis}, {Maiolino}, {Nordon}, {P{\'e}rez Fournon},
  {P{\'e}rez Garc{\'{\i}}a}, {Poglitsch}, {Popesso}, {Pozzi}, {Riguccini},
  {Rodighiero}, {Saintonge}, {Santini}, {Sanchez-Portal}, {Shao}, {Sturm},
  {Tacconi}, {Valtchanov}, {Wieprecht}, \& {Wiezorrek}}]{Magnelli:2010}
{Magnelli}, B., {et~al.} 2010, \aap, 518, L28

\bibitem[{Matteo {et~al.}(2005)Matteo, Springel, \& Hernquist}]{DiMatteo:2005}
Matteo, T.~D., Springel, V., \& Hernquist, L. 2005, \nat, 433, 604

\bibitem[{{Men{\'e}ndez-Delmestre} {et~al.}(2007)}]{Menendez:2007}
{Men{\'e}ndez-Delmestre}, K., {et~al.} 2007, \apjl, 655, L65

\bibitem[{{Men{\'e}ndez-Delmestre} {et~al.}(2009)}]{Menendez:2009}
---. 2009, \apj, 699, 667

\bibitem[{{Micha{\l}owski} {et~al.}(2010{\natexlab{a}}){Micha{\l}owski},
  {Hjorth}, \& {Watson}}]{Michalowski:2010masses}
{Micha{\l}owski}, M., {Hjorth}, J., \& {Watson}, D. 2010{\natexlab{a}}, \aap,
  514, A67

\bibitem[{{Micha{\l}owski} {et~al.}(2010{\natexlab{b}}){Micha{\l}owski},
  {Watson}, \& {Hjorth}}]{Michalowski:2010production}
{Micha{\l}owski}, M.~J., {Watson}, D., \& {Hjorth}, J. 2010{\natexlab{b}},
  \apj, 712, 942

\bibitem[{{Mihos} \& {Hernquist}(1994)}]{Mihos:1994}
{Mihos}, J.~C., \& {Hernquist}, L. 1994, \apjl, 437, L47

\bibitem[{{Mihos} \& {Hernquist}(1996)}]{Mihos:1996}
---. 1996, \apj, 464, 641

\bibitem[{Narayanan {et~al.}(2011)Narayanan, {Cox}, {Hayward}, \&
  {Hernquist}}]{Narayanan:2011ks}
Narayanan, D., {Cox}, T.~J., {Hayward}, C.~C., \& {Hernquist}, L. 2011, \mnras,
  412, 287

\bibitem[{Narayanan {et~al.}(2009)Narayanan, Cox, Hayward, Younger, \&
  Hernquist}]{Narayanan:2009}
Narayanan, D., Cox, T.~J., Hayward, C.~C., Younger, J.~D., \& Hernquist, L.
  2009, \mnras, 400, 1919

\bibitem[{Narayanan {et~al.}(2008{\natexlab{a}})Narayanan, Cox, Shirley,
  Dav{\'e}, Hernquist, \& Walker}]{Narayanan:2008CO_SFR}
Narayanan, D., Cox, T.~J., Shirley, Y., Dav{\'e}, R., Hernquist, L., \& Walker,
  C.~K. 2008{\natexlab{a}}, \apj, 684, 996

\bibitem[{Narayanan {et~al.}(2010{\natexlab{a}})Narayanan, Hayward, Cox,
  Hernquist, Jonsson, Younger, \& Groves}]{Narayanan:2010smg}
Narayanan, D., Hayward, C.~C., Cox, T.~J., Hernquist, L., Jonsson, P., Younger,
  J.~D., \& Groves, B. 2010{\natexlab{a}}, \mnras, 401, 1613 (N10)

\bibitem[{Narayanan {et~al.}(2008{\natexlab{b}})Narayanan, {Li}, {Cox},
  {Hernquist}, {Hopkins}, {Chakrabarti}, {Dav{\'e}}, {Di Matteo}, {Gao},
  {Kulesa}, {Robertson}, \& {Walker}}]{Narayanan:2008quasar_CO}
Narayanan, D., {et~al.} 2008{\natexlab{b}}, \apjs, 174, 13

\bibitem[{Narayanan {et~al.}(2010{\natexlab{b}})Narayanan, {Dey}, {Hayward},
  {Cox}, {Bussmann}, {Brodwin}, {Jonsson}, {Hopkins}, {Groves}, {Younger}, \&
  {Hernquist}}]{Narayanan:2010dog}
---. 2010{\natexlab{b}}, \mnras, 407, 1701

\bibitem[{Neri {et~al.}(2003)Neri, Genzel, Ivison, Bertoldi, Blain, Chapman,
  Cox, Greve, Omont, \& Frayer}]{Neri:2003}
Neri, R., {et~al.} 2003, \apj, 597, L113

\bibitem[{Noeske {et~al.}(2007{\natexlab{a}})Noeske, Faber, Weiner, Koo,
  Primack, Dekel, Papovich, Conselice, Floc'h, Rieke, Coil, Lotz, Somerville,
  \& Bundy}]{Noeske:2007b}
Noeske, K.~G., {et~al.} 2007{\natexlab{a}}, \apjl, 660, L47

\bibitem[{Noeske {et~al.}(2007{\natexlab{b}})Noeske, Weiner, Faber, Papovich,
  Koo, Somerville, Bundy, Conselice, Newman, Schiminovich, Floc'h, Coil, Rieke,
  Lotz, Primack, Barmby, Cooper, Davis, Ellis, Fazio, Guhathakurta, Huang,
  Kassin, Martin, Phillips, Rich, Small, Willmer, \& Wilson}]{Noeske:2007a}
---. 2007{\natexlab{b}}, \apjl, 660, L43

\bibitem[{Papadopoulos {et~al.}(2010)Papadopoulos, van~der Werf, Isaak, \&
  Xilouris}]{Papadopoulos:2010}
Papadopoulos, P.~P., van~der Werf, P., Isaak, K., \& Xilouris, E.~M. 2010,
  \apj, 715, 775

\bibitem[{Peacock {et~al.}(2000)Peacock, Rowan-Robinson, Blain, Dunlop,
  Efstathiou, Hughes, Jenness, Ivison, Lawrence, Longair, Mann, Oliver, \&
  Serjeant}]{Peacock:2000}
Peacock, J.~A., {et~al.} 2000, \mnras, 318, 535

\bibitem[{{Pope} {et~al.}(2006){Pope}, {Scott}, {Dickinson}, {Chary},
  {Morrison}, {Borys}, {Sajina}, {Alexander}, {Daddi}, {Frayer}, {MacDonald},
  \& {Stern}}]{Pope:2006}
{Pope}, A., {et~al.} 2006, \mnras, 370

\bibitem[{{Pope} {et~al.}(2008)}]{Pope:2008MIR}
---. 2008, \apj, 675, 1171

\bibitem[{{Rangwala} {et~al.}(2011){Rangwala}, {Maloney}, {Glenn}, {Wilson},
  {Rykala}, {Isaak}, {Baes}, {Bendo}, {Boselli}, {Bradford}, {Clements},
  {Cooray}, {Fulton}, {Imhof}, {Kamenetzky}, {Madden}, {Mentuch}, {Sacchi},
  {Sauvage}, {Schirm}, {Smith}, {Spinoglio}, \& {Wolfire}}]{Rangwala:2011}
{Rangwala}, N., {et~al.} 2011, arXiv:1106.5054

\bibitem[{{Rex} {et~al.}(2010){Rex}, {Rawle}, {Egami},
  {P{\'e}rez-Gonz{\'a}lez}, {Zemcov}, {Aretxaga}, {Chung}, {Fadda}, {Gonzalez},
  {Hughes}, {Horellou}, {Johansson}, {Kneib}, {Richard}, {Altieri}, {Fiedler},
  {Pereira}, {Rieke}, {Smail}, {Valtchanov}, {Blain}, {Bock}, {Boone},
  {Bridge}, {Clement}, {Combes}, {Dowell}, {Dessauges-Zavadsky}, {Ilbert},
  {Ivison}, {Jauzac}, {Lutz}, {Omont}, {Pell{\'o}}, {Rodighiero}, {Schaerer},
  {Smith}, {Walth}, {van der Werf}, {Werner}, {Austermann}, {Ezawa}, {Kawabe},
  {Kohno}, {Perera}, {Scott}, {Wilson}, \& {Yun}}]{Rex:2010}
{Rex}, M., {et~al.} 2010, \aap, 518, L13

\bibitem[{{Ricciardelli} {et~al.}(2010){Ricciardelli}, {Trujillo}, {Buitrago},
  \& {Conselice}}]{Ricciardelli:2010}
{Ricciardelli}, E., {Trujillo}, I., {Buitrago}, F., \& {Conselice}, C.~J. 2010,
  \mnras, 406, 230

\bibitem[{{Robertson} {et~al.}(2006{\natexlab{a}}){Robertson}, {Bullock},
  {Cox}, {Di Matteo}, {Hernquist}, {Springel}, \&
  {Yoshida}}]{Robertson:2006disk_formation}
{Robertson}, B., {Bullock}, J.~S., {Cox}, T.~J., {Di Matteo}, T., {Hernquist},
  L., {Springel}, V., \& {Yoshida}, N. 2006{\natexlab{a}}, \apj, 645, 986

\bibitem[{{Robertson} {et~al.}(2006{\natexlab{b}}){Robertson}, Hernquist, Cox,
  Matteo, Hopkins, Martini, \& Springel}]{Robertson:2006}
{Robertson}, B., Hernquist, L., Cox, T.~J., Matteo, T.~D., Hopkins, P.~F.,
  Martini, P., \& Springel, V. 2006{\natexlab{b}}, \apj, 641, 90

\bibitem[{{Robertson} \& {Bullock}(2008)}]{Robertson:2008}
{Robertson}, B.~E., \& {Bullock}, J.~S. 2008, \apjl, 685, L27

\bibitem[{{Rothberg} \& {Fischer}(2010)}]{Rothberg:2010}
{Rothberg}, B., \& {Fischer}, J. 2010, \apj, 712, 318

\bibitem[{{Rothberg} \& {Joseph}(2004)}]{Rothberg:2004}
{Rothberg}, B., \& {Joseph}, R.~D. 2004, \aj, 128, 2098

\bibitem[{Sakamoto {et~al.}(1999)Sakamoto, Scoville, Yun, Crosas, Genzel, \&
  Tacconi}]{Sakamoto:1999}
Sakamoto, K., Scoville, N.~Z., Yun, M.~S., Crosas, M., Genzel, R., \& Tacconi,
  L.~J. 1999, \apj, 514, 68

\bibitem[{Sakamoto {et~al.}(2008)Sakamoto, Wang, Wiedner, Wang, Peck, Zhang,
  Petitpas, Ho, \& Wilner}]{Sakamoto:2008}
Sakamoto, K., {et~al.} 2008, \apj, 684, 957

\bibitem[{{Sanders} \& {Mirabel}(1996)}]{Sanders:1996}
{Sanders}, D.~B., \& {Mirabel}, I.~F. 1996, \araa, 34, 749

\bibitem[{Schinnerer {et~al.}(2008)Schinnerer, Carilli, Capak,
  Martinez-Sansigre, Scoville, Smol{\v c}i{\'c}, Taniguchi, Yun, Bertoldi,
  Fevre, \& de~Ravel}]{Schinnerer:2008}
Schinnerer, E., {et~al.} 2008, \apjl, 689, L5

\bibitem[{{Scott} {et~al.}(2002){Scott}, {Fox}, {Dunlop}, {Serjeant},
  {Peacock}, {Ivison}, {Oliver}, {Mann}, {Lawrence}, {Efstathiou},
  {Rowan-Robinson}, {Hughes}, {Archibald}, {Blain}, \& {Longair}}]{Scott:2002}
{Scott}, S.~E., {et~al.} 2002, \mnras, 331, 817

\bibitem[{Scoville {et~al.}(1991)Scoville, Sargent, Sanders, \&
  Soifer}]{Scoville:1991}
Scoville, N.~Z., Sargent, A.~I., Sanders, D.~B., \& Soifer, B.~T. 1991, \apj,
  366, L5

\bibitem[{{Shetty} {et~al.}(2009{\natexlab{a}}){Shetty}, {Kauffmann}, {Schnee},
  \& {Goodman}}]{Shetty:2009a}
{Shetty}, R., {Kauffmann}, J., {Schnee}, S., \& {Goodman}, A.~A.
  2009{\natexlab{a}}, \apj, 696, 676

\bibitem[{{Shetty} {et~al.}(2009{\natexlab{b}}){Shetty}, {Kauffmann}, {Schnee},
  {Goodman}, \& {Ercolano}}]{Shetty:2009b}
{Shetty}, R., {Kauffmann}, J., {Schnee}, S., {Goodman}, A.~A., \& {Ercolano},
  B. 2009{\natexlab{b}}, \apj, 696, 2234

\bibitem[{Smail {et~al.}(2004)Smail, Chapman, Blain, \& Ivison}]{Smail:2004}
Smail, I., Chapman, S.~C., Blain, A.~W., \& Ivison, R.~J. 2004, \apj, 616, 71

\bibitem[{Smail {et~al.}(1997)Smail, Ivison, \& Blain}]{Smail:1997}
Smail, I., Ivison, R.~J., \& Blain, A.~W. 1997, \apjl, 490, L5

\bibitem[{{Snyder} {et~al.}(2011){Snyder}, {Cox}, {Hayward}, {Hernquist}, \&
  {Jonsson}}]{Snyder:2011}
{Snyder}, G.~F., {Cox}, T.~J., {Hayward}, C.~C., {Hernquist}, L., \& {Jonsson},
  P. 2011, ArXiv e-prints, arXiv:1102.3689

\bibitem[{{Somerville} {et~al.}(2008){Somerville}, {Hopkins}, {Cox},
  {Robertson}, \& {Hernquist}}]{Somerville:2008}
{Somerville}, R.~S., {Hopkins}, P.~F., {Cox}, T.~J., {Robertson}, B.~E., \&
  {Hernquist}, L. 2008, \mnras, 391, 481

\bibitem[{Springel(2005)}]{Springel:2005gadget}
Springel, V. 2005, \mnras, 364, 1105

\bibitem[{Springel \& Hernquist(2002)}]{Springel:2002}
Springel, V., \& Hernquist, L. 2002, \mnras, 333, 649

\bibitem[{Springel \& Hernquist(2003)}]{Springel:2003}
---. 2003, \mnras, 339, 289

\bibitem[{Springel {et~al.}(2005)Springel, Matteo, \&
  Hernquist}]{Springel:2005feedback}
Springel, V., Matteo, T.~D., \& Hernquist, L. 2005, \mnras, 361, 776

\bibitem[{{Springel} {et~al.}(2001){Springel}, {Yoshida}, \&
  {White}}]{Springel:2001gadget}
{Springel}, V., {Yoshida}, N., \& {White}, S.~D.~M. 2001, NewA, 6, 79

\bibitem[{Swinbank {et~al.}(2004)Swinbank, Smail, Chapman, Blain, Ivison, \&
  Keel}]{Swinbank:2004}
Swinbank, A.~M., Smail, I., Chapman, S.~C., Blain, A.~W., Ivison, R.~J., \&
  Keel, W.~C. 2004, \apj, 617, 64

\bibitem[{Swinbank {et~al.}(2008)Swinbank, Lacey, Smail, Baugh, Frenk, Blain,
  Chapman, Coppin, Ivison, Gonzalez, \& Hainline}]{Swinbank:2008}
Swinbank, A.~M., {et~al.} 2008, \mnras, 391, 420

\bibitem[{Tacconi {et~al.}(2006)Tacconi, Neri, Chapman, Genzel, Smail, Ivison,
  Bertoldi, Blain, Cox, Greve, \& Omont}]{Tacconi:2006}
Tacconi, L.~J., {et~al.} 2006, \apj, 640, 228

\bibitem[{Tacconi {et~al.}(2008)Tacconi, Genzel, Smail, Neri, Chapman, Ivison,
  Blain, Cox, Omont, Bertoldi, Greve, Schreiber, Genel, Lutz, Swinbank,
  Shapley, Erb, Cimatti, Daddi, \& Baker}]{Tacconi:2008}
---. 2008, \apj, 680, 246

\bibitem[{Tacconi {et~al.}(2010)Tacconi, Genzel, Neri, Cox, Cooper, Shapiro,
  Bolatto, Bouch{\'e}, Bournaud, Burkert, Combes, Comerford, Davis,
  {F{\"o}rster Schreiber}, Garcia-Burillo, Gracia-Carpio, Lutz, Naab, Omont,
  Shapley, Sternberg, \& Weiner}]{Tacconi:2010}
---. 2010, \nat, 463, 781

\bibitem[{{Targett} {et~al.}(2011){Targett}, {Dunlop}, {McLure}, {Best},
  {Cirasuolo}, \& {Almaini}}]{Targett:2011}
{Targett}, T.~A., {Dunlop}, J.~S., {McLure}, R.~J., {Best}, P.~N., {Cirasuolo},
  M., \& {Almaini}, O. 2011, \mnras, in press, arXiv:1005.5176

\bibitem[{{Valiante} {et~al.}(2007)}]{Valiante:2007}
{Valiante}, E., {et~al.} 2007, \apj, 660, 1060

\bibitem[{van Kampen {et~al.}(2005)van Kampen, Percival, Crawford, Dunlop,
  Scott, Bevis, Oliver, Pearce, Kay, Gazta{\~n}aga, Hughes, \&
  Aretxaga}]{vanKampen:2005}
van Kampen, E., {et~al.} 2005, \mnras, 359, 469

\bibitem[{V{\'a}rosi \& Dwek(1999)}]{Varosi:1999a}
V{\'a}rosi, F., \& Dwek, E. 1999, \apj, 523, 265

\bibitem[{{Wang} {et~al.}(2011){Wang}, {Cowie}, {Barger}, \&
  {Williams}}]{Wang:2011}
{Wang}, W.-H., {Cowie}, L.~L., {Barger}, A.~J., \& {Williams}, J.~P. 2011,
  \apjl, 726, L18+

\bibitem[{Webb {et~al.}(2003)Webb, Eales, Foucaud, Lilly, McCracken,
  Adelberger, Steidel, Shapley, Clements, Dunne, F{\`e}vre, Brodwin, \&
  Gear}]{Webb:2003}
Webb, T.~M., {et~al.} 2003, \apj, 582, 6

\bibitem[{Weingartner \& Draine(2001)}]{Weingartner:2001}
Weingartner, J.~C., \& Draine, B.~T. 2001, \apj, 548, 296

\bibitem[{Wilson {et~al.}(2008)Wilson, Austermann, Perera, Scott, Ade, Bock,
  Glenn, Golwala, Kim, Kang, Lydon, Mauskopf, Predmore, Roberts, Souccar, \&
  Yun}]{Wilson:2008}
Wilson, G.~W., {et~al.} 2008, \mnras, 386, 807

\bibitem[{Witt \& Gordon(1996)}]{Witt:1996}
Witt, A.~N., \& Gordon, K.~D. 1996, \apj, 463, 681

\bibitem[{{Wright}(2006)}]{Wright:2006}
{Wright}, E.~L. 2006, \pasp, 118, 1711

\bibitem[{Wuyts {et~al.}(2010)Wuyts, Cox, Hayward, Franx, Hernquist, Hopkins,
  Jonsson, \& van Dokkum}]{Wuyts:2010}
Wuyts, S., Cox, T.~J., Hayward, C.~C., Franx, M., Hernquist, L., Hopkins,
  P.~F., Jonsson, P., \& van Dokkum, P.~G. 2010, \apj, 722, 1666

\bibitem[{Wuyts {et~al.}(2009)Wuyts, Franx, Cox, {F{\"o}rster Schreiber},
  Hayward, Hernquist, Hopkins, Labb{\'e}, Marchesini, Robertson, Toft, \& van
  Dokkum}]{Wuyts:2009b}
Wuyts, S., {et~al.} 2009, \apj, 700, 799

\bibitem[{{Younger} {et~al.}(2009{\natexlab{a}}){Younger}, Hayward, Narayanan,
  Cox, Hernquist, \& Jonsson}]{Younger:2009}
{Younger}, J.~D., Hayward, C.~C., Narayanan, D., Cox, T.~J., Hernquist, L., \&
  Jonsson, P. 2009{\natexlab{a}}, \mnras, 396, L66

\bibitem[{{Younger} {et~al.}(2007){Younger}, {Fazio}, {Huang}, {Yun}, {Wilson},
  {Ashby}, {Gurwell}, {Lai}, {Peck}, {Petitpas}, {Wilner}, {Iono}, {Kohno},
  {Kawabe}, {Hughes}, {Aretxaga}, {Webb}, {Mart{\'{\i}}nez-Sansigre}, {Kim},
  {Scott}, {Austermann}, {Perera}, {Lowenthal}, {Schinnerer}, \& {Smol{\v
  c}i{\'c}}}]{Younger:2007high-z_SMGs}
{Younger}, J.~D., {et~al.} 2007, \apj, 671, 1531

\bibitem[{{Younger} {et~al.}(2008){Younger}, {Fazio}, {Wilner}, {Ashby},
  {Blundell}, {Gurwell}, {Huang}, {Iono}, {Peck}, {Petitpas}, {Scott},
  {Wilson}, \& {Yun}}]{Younger:2008phys_scale}
---. 2008, \apj, 688, 59

\bibitem[{{Younger} {et~al.}(2009{\natexlab{b}}){Younger}, {Omont}, {Fiolet},
  {Huang}, {Fazio}, {Lai}, {Polletta}, {Rigopoulou}, \&
  {Zylka}}]{Younger:2009EGS}
---. 2009{\natexlab{b}}, \mnras, 394, 1685

\bibitem[{{Younger} {et~al.}(2009{\natexlab{c}}){Younger}, {Fazio}, {Huang},
  {Yun}, {Wilson}, {Ashby}, {Gurwell}, {Peck}, {Petitpas}, {Wilner}, {Hughes},
  {Aretxaga}, {Kim}, {Scott}, {Austermann}, {Perera}, \&
  {Lowenthal}}]{Younger:2009SMG_interf}
---. 2009{\natexlab{c}}, \apj, 704, 803

\bibitem[{{Younger} {et~al.}(2010){Younger}, {Fazio}, {Ashby}, {Civano},
  {Gurwell}, {Huang}, {Iono}, {Peck}, {Petitpas}, {Scott}, {Wilner}, {Wilson},
  \& {Yun}}]{Younger:2010}
---. 2010, \mnras, 407, 1268

\end{thebibliography}

\appendix

\section{Derivation of the relations given in \S3.2.1}

Here we derive the relations for submm flux density as a function of dust bolometric luminosity $\ldust$ and dust mass $\mdust$ (Equation \ref{eq:single_T_L_scaling})
and SFR and $\mdust$ (Equation \ref{eq:single_T_SFR_scaling}) for an optically thin modified blackbody. One can model galaxy SEDs with more complex models
\citep[e.g.,][]{Dale:2001,Dale:2002,Chakrabarti:2005RT,Chakrabarti:2008RT,Kovacs:2010}, but for the sake of simplicity and because the optically thin modified
blackbody is commonly used for SED fitting we only consider an optically thin modified blackbody here.
Consider a mass $\mdust$ of dust with temperature $\tdust$.
Assuming the dust is optically thin at rest-frame frequency $\nu_r$, the luminosity density emitted by the dust at that frequency is
\begin{equation} \label{eq:mod_BB_L_nu}
L_{\nu_r} = 4 \pi \kappa_{\nu_r} \mdust B_{\nu_r} (\tdust),
\end{equation}
where $\kappa_{\nu_r}$ is the dust opacity (m$^2$ kg$^{-1}$) at rest-frame frequency $\nu_r$ and $B_{\nu_r}(\tdust)$ is the Planck function.
We assume a power-law opacity in the IR,
\begin{equation}
\kappa_{\nu_r} = \kappa_0 \left(\frac{\nu_r}{\nu_0}\right)^{\beta},
\end{equation}
where $\kappa_0$ is the opacity at frequency $\nu_0$.
Integrating Equation (\ref{eq:mod_BB_L_nu}) over $\nu$ gives the total dust luminosity,
\begin{equation} \label{eq:L_dust}
\ldust = \Gamma(4+\beta) \zeta(4+\beta) \frac{8 \pi h}{c^2} \left(\frac{k \tdust}{h}\right)^4 \mdust \kappa_0 \left(\frac{k \tdust}{h \nu_0}\right)^{\beta},
\end{equation}
where $\Gamma$ and $\zeta$ are Riemann functions, $h$ is the Planck constant, $c$ is the speed of light, and $k$ is the Boltzmann constant.

Solving for the effective dust temperature yields
\begin{equation} \label{eq:T_dust}
T_d = \frac{h}{k} \left[\frac{L_d c^2 \nu_0^{\beta}}{\Gamma(4+\beta) \zeta(4+\beta) 8 \pi \kappa_0 h \mdust}\right]^{1/(4+\beta)}.
\end{equation}

If we place the mass of dust at redshift $z$, the flux density at observed-frame frequency $\nu_o$ is
\begin{eqnarray}
S_{\nu_o}(\tdust) &=& (1+z) \frac{L_{\nu_r}}{4 \pi D_L^2} \\
&=& (1+z) \frac{4 \pi \kappa_{\nu_r} \mdust B_{\nu_r} (\tdust)}{4 \pi D_L^2} \\
&=& (1+z)^{\beta-3} \frac{\kappa_0 \mdust}{D_A^2} \left( \frac{\nu_o}{\nu_0} \right)^{\beta} B_{\nu_o(1+z)}(\tdust),
\end{eqnarray}
where we have related angular diameter distance $D_A$ and luminosity distance $D_L$ using $D_L = (1+z)^2 D_A.$
In the Rayleigh-Jeans limit, $B_{\nu}(T) = 2k \nu^2 T/c^2$, so
\begin{equation} \label{eq:mod_BB_S_nu_RJ}
S_{\nu_o}(\tdust) = (1+z)^{\beta - 1} \frac{2 k \kappa_0}{c^2 D_A^2} \left( \frac{\nu_o}{\nu_0} \right)^{\beta} \nu_o^2 \mdust \tdust.
\end{equation}
By substituting $T_d$ from Equation (\ref{eq:T_dust}) into Equation (\ref{eq:mod_BB_S_nu_RJ}) we find
\begin{equation} \label{eq:S_nu_general}
S_{\nu_o} = \frac{2h\kappa_0}{c^2 D_A^2} \nu_o^2 \left(\frac{\nu_o}{\nu_0}\right)^{\beta} \left(\frac{\nu_0^{\beta} c^2}{\Gamma(4+\beta) \zeta(4+\beta)
8 \pi \kappa_0 h} \right)^{1/(4+\beta)} (1+z)^{\beta-1} \ldust^{1/(4+\beta)} \mdust^{(3+\beta)/(4+\beta)}.
\end{equation}

For the \citet{Weingartner:2001} $R_V = 3.1$ Milky Way dust model, which we use in our RT
calculations, $\beta \approx 2$ and the
$850$ \micron ~opacity is $\kappa_{850} = 0.050$ m$^2$ kg$^{-1}$, consistent
with the value \citet{James:2002} derived from submm observations of local galaxies, $0.07 \pm 0.02$ m$^2$ kg$^{-1}$,
and with the results of \citet{Dunne:2003}.
Thus the observed 850 \micron ~flux density is
\begin{equation}
S_{850} = 1.5 {\rm ~mJy} ~(1+z) \left( \frac{D_A}{1 {\rm ~Gpc}} \right)^{-2} \left(\frac{\ldust}{10^{12} \lsun}\right)^{1/6} \left(\frac{\mdust}{10^8 \msun}\right)^{5/6}.
\end{equation}

The \citet{Kennicutt:1998review} SFR-$\lir$ calibration converted to a \citet{Kroupa:2001} initial mass function
is
\begin{equation} \label{eq:K98_IR_SFR_cal}
\lir \approx \lbol \approx 9 \times 10^9 \lsun ({\rm SFR}/\msunperyr).
\end{equation}
This conversion assumes all starlight is absorbed by dust and the contribution from AGN and old stars is negligible; as discussed above,
these assumptions are all violated at some level. (If these assumption were true, the power-law indices in Equations (\ref{eq:sfr_fitting_functions}) and
(\ref{eq:L_fitting_functions}) would be identical.)
However, since the above calibration is ubiquitously applied, we will give the relation that results when we use it; the relation is
\begin{equation}
S_{850} = 1.5 {\rm ~mJy} ~(1+z) \left( \frac{D_A}{1 {\rm ~Gpc}} \right)^{-2} \left(\frac{\rm ~SFR}{100 ~\msunperyr}\right)^{1/6}
\left(\frac{\mdust}{10^8 \msun}\right)^{5/6}.
\end{equation}

Assuming $\Omega_m = 0.27, \Omega_{\Lambda} = 0.73$, and $h = 0.7$, the angular diameter distance at $z = 2$ is 1.77 Gpc \citep{Wright:2006}.
Thus for $z = 2$ we recover Equation (\ref{eq:single_T_L_scaling}),
\begin{equation}
S_{850} = 1.4 {\rm ~mJy} \left(\frac{\ldust}{10^{12} \lsun}\right)^{1/6} \left(\frac{\mdust}{10^8 \msun}\right)^{5/6}.
\end{equation}
This should be compared to Equation (\ref{eq:L_fitting_functions}). In terms of SFR, we get Equation (\ref{eq:single_T_SFR_scaling}),
\begin{equation}
S_{850} = 1.4 {\rm ~mJy} \left(\frac{\rm ~SFR}{100 ~\msunperyr}\right)^{1/6} \left(\frac{\mdust}{10^8 \msun}\right)^{5/6}.
\end{equation}
This should be compared to Equation (\ref{eq:sfr_fitting_functions}).

Even if the underlying power-law index of the dust opacity curve is $\beta = 2$, for a distribution of dust temperatures a single-temperature modified blackbody with
$\beta = 1.5$ may better fit the SED \citep{Dunne:2001,Chakrabarti:2008RT}. Note also that the fitted dust temperature and $\beta$ are degenerate, and the fitted
values can depend sensitively on both noise in the data and temperature variations along the line-of-sight \citep{Shetty:2009a,Shetty:2009b}.
Since $\beta = 1.5$ is often assumed when fitting SEDs and determining dust
masses of SMGs \citep[e.g.,][]{Kovacs:2006,Kovacs:2010,Coppin:2008,Chapman:2010}, so we will provide the relations for $\beta = 1.5$ also. They are:
\begin{equation}
S_{850} = 1.9 {\rm ~mJy} ~(1+z)^{0.5} \left( \frac{D_A}{1 {\rm ~Gpc}} \right)^{-2} \left(\frac{\ldust}{10^{12} \lsun}\right)^{0.18} \left(\frac{\mdust}{10^8 \msun}\right)^{0.82},
\end{equation}
and
\begin{equation}
S_{850} = 1.9 {\rm ~mJy} ~(1+z)^{0.5} \left( \frac{D_A}{1 {\rm ~Gpc}} \right)^{-2} \left(\frac{\rm ~SFR}{100 ~\msunperyr}\right)^{0.18} \left(\frac{\mdust}{10^8 \msun}\right)^{0.82}.
\end{equation}

For $z = 2$,
\begin{equation}
S_{850} = 1.0 {\rm ~mJy} \left(\frac{\ldust}{10^{12} \lsun}\right)^{0.18} \left(\frac{\mdust}{10^8 \msun}\right)^{0.82},
\end{equation}
and
\begin{equation}
S_{850} = 1.0 {\rm ~mJy} \left(\frac{\rm ~SFR}{100 ~\msunperyr}\right)^{0.18} \left(\frac{\mdust}{10^8 \msun}\right)^{0.82}.
\end{equation}

The equations can be rescaled to different values of $\kappa_0$ using $S_{\nu_o} \propto \kappa_0^{(3+\beta)/(4+\beta)}$
and to different submm wavelengths using $S_{\nu_o} \propto \nu_o^{2+\beta}$ (see Equation \ref{eq:S_nu_general}).

\end{document}